\documentclass{ar2e}
\usepackage{ulem}  
\usepackage{ARAstroBib,natbib}
\usepackage{color}
\begin{document}
\input epsf.tex    

\input psfig.sty

\jname{Annu. Rev. Astron. Astrophys.}
\jyear{2013}
\jvol{51}
\ARinfo{1056-8700/97/0610-00}


\title{Solar Irradiance Variability and Climate}

\markboth{S.K.~Solanki, N.A.~Krivova, J.D.~Haigh}{Solar Activity and Climate}

\author{
 Sami~K.~Solanki, 
\affiliation{Max-Planck-Institut f\"ur Sonnensystemforschung, 37191
 Katlenburg-Lindau, Germany}
\affiliation{School of Space Research, Kyung Hee University, Yongin,
 Gyeonggi 446-701, Korea}
 Natalie~A.~Krivova
\affiliation{Max-Planck-Institut f\"ur Sonnensystemforschung, 37191
 Katlenburg-Lindau, Germany}
 Joanna~D.~Haigh
\affiliation{Imperial College, London SW7 2AZ, U.K.}
}


\begin{abstract}
The brightness of the Sun varies on all time scales on which it has been
observed, and there is increasing evidence that it has an influence on
climate.
The amplitudes of such variations depend on the wavelength and possibly on
the time scale.
Although many aspects of this variability are well established, the exact
magnitude of secular variations (going beyond a solar cycle) and the
spectral dependence of variations are under discussion.
The main drivers of solar variability are thought to be magnetic features at
the solar surface.
The climate reponse can be, on a global scale, largely accounted for by
simple energetic considerations, but understanding the regional climate
effects is more difficult.
Promising mechanisms for such a driving have been identified, including
through the influence of UV irradiance on the stratosphere and dynamical
coupling to the surface. Here we provide an overview of the current state of
our knowledge, as well as of the main open questions.
\end{abstract}

\maketitle


\section{INTRODUCTION}
\label{intro}

The Sun is a very special star. Not only is it a boon to astronomers in that
it allows us to start resolving spatial scales at which universal physical
processes take place that act also in other astronomical objects.  It is
also the only star that directly influences the Earth and thus also our
lives.

Of the many ways in which the Sun affects the Earth, the most obvious is
by its radiation.  The approximately 1361 Wm$^{-2}$ received from the
present day Sun at 1 AU (the total solar irradiance, see below) are
responsible for keeping the Earth from cooling off to temperatures that are
too low for sustaining human life.  The composition, structure and dynamics
of the Earth's atmosphere also play a very fundamental part by making
efficient use (through the greenhouse effect) of the energy input from the
Sun.

The total solar irradiance, or TSI, is defined as the total power from
the Sun impinging on a unit area (perpendicular to the Sun's rays) at 1AU
(given in units of Wm$^{-2}$).
The TSI is the wavelength integral over the solar spectral irradiance, or SSI
(Wm$^{-2}$nm$^{-1}$).

Under normal circumstances,
the Sun is the only serious external source of energy to Earth. Any
variability of the Sun's radiative output thus has the potential of
affecting our climate and hence the habitability of the Earth.  The
important question is how strong this influence is and in particular how it
compares with other mechanisms including the influence of man-made
greenhouse gases. 
Although this has been debated for a long time, the debate is being
held with increasing urgency due to the unusual global temperature rise we
have seen in the course of the 20th century and particularly during the last
3--4 decades.
It is generally agreed that the recent warming is mainly driven by the
release of greenhouse gases, foremost among them carbon dioxide, into the
Earth's atmosphere by the burning of fossil fuels \citep{ipcc-2007}. 
However, determining the exact level of warming due to man-made greenhouse
gases requires a good understanding of the natural causes of climate change.
These natural causes are partly to be found in the climate system itself
(which includes the oceans and the land surfaces), partly they come from the
Earth's interior, by the release of aerosols and dust through volcanoes, and
partly they lie outside the Earth and are thus astronomical in nature.

A variety of astronomical effects can influence the Earth's
climate.  Thus the energetic radiation from a nearby supernova could
adversely affect our atmosphere in a dramatic fashion
\citep[e.g.,][]{Svensmark2012}.
Also, modulation of cosmic rays as the Sun passes into and out of spiral
arms during its orbit around the galaxy has been proposed to explain slow
variations in climate taking place over 100s of millions of years
\citep{Shaviv2002}.

However, the most obvious astronomical influence is due, directly or
indirectly, to the Sun, which is the source of practically all external
energy input into the climate system.
The Sun's influence can follow three different paths:
1) variations in insolation through changes in the Sun's radiative output
itself (direct influence);
2) modulations of the radiation reaching different hemispheres of the Earth
through changes in the Earth's orbital parameters and in the obliquity of its
rotation axis (indirect influence);
3) the influence of the Sun's activity on galactic cosmic rays proposed to
affect cloud cover by, e.g., \citet{Marsh2000}.

The first of these is generally considered to be the main cause of the
solar contribution to global climate change and will be described in greater detail
below. We need to distinguish between changes in TSI, i.e. in the
energy input to the Earth system, and variations in SSI, particularly in UV
irradiance, which can enhance the Sun's effect by impacting on the
chemistry in the Earth's middle atmosphere.

The second path is now accepted as the prime cause of the pattern of ice
ages and the interglacial warm periods that have dominated the longer term
evolution of the climate over the past few million years.  The various
parameters of the Earth's orbital and rotational motion vary at periods of
23~kyr (precession), 41~kyr (obliquity) and 100~kyr (eccentricity)
\citep[e.g.,][]{Paillard2001,Crucifix2006}.  The changes in the Earth's
orbit are so slow that they are unlikely to have contributed to the global
warming over the last century.

The third potential path builds on the modulation of the flux of galactic cosmic
rays by solar magnetic activity.  The Sun's open magnetic flux
(i.e.  the flux in the field lines reaching out into the heliosphere) and
the solar wind impede the propagation of the charged galactic cosmic rays
into the inner solar system, so that at times of high solar activity fewer
cosmic rays reach Earth.  Their connection with climate has been drawn by,
e.g., \citet{Marsh2000} from the correlation between the cosmic ray flux and
global cloud cover.  However, this mechanism still has to establish itself. 
Thus the CLOUD experiment at CERN has so far returned only equivocal results
on the effectiveness of cosmic rays in producing clouds \citep{Kirkby2011}.

In this review we provide an overview of our knowledge of solar irradiance
variability and of the response of the Earth's climate to changes in solar
irradiance. 
Consequently, mechanisms 2 and 3 are not considered further here.
A number of earlier reviews have also covered  all or aspects of this topic.
Thus, overviews of solar irradiance variability have been given by
\citet{Lean1997}, \citet{froehlich-lean-2004}, \citet{Solanki2005},
\citet{domingo-et-al-2009}, \citet{lean-deland-2012},
\citet{krivova-solanki-2012}.
The solar activity variations underlying irradiance changes have been
reviewed by \citet{Usoskin2008}, \citet{hathaway-2010},
\citet{Charbonneau2010}, \citet{Usoskin2012}.
Solar irradiance together with the response of the Earth's atmosphere have
been covered by \citet{Haigh2005}, \citet{Haigh2007}, \citet{Gray2010},
\citet{ermolli-et-al-2012a}, cf.  the monograph edited by \citet{Pap2004}.

In the following we first discuss the measurements of solar irradiance
variations, their causes and models aiming to reproduce the data (Sect.~2),
followed by an overview of
longer term evolution of solar activity and the associated evolution of solar
irradiance (Sect.~3). In Sect.~4 we then move to the response of the Earth's
atmosphere to solar irradiance variations, with conclusions being given in
Sect.~5.



\section{SHORT-TERM SOLAR IRRADIANCE VARIABILITY}
\label{short-term}

\subsection{Measurements of TSI and SSI}
\label{obs}

The measurement of solar irradiance with an accuracy sufficiently
high to detect and reliably follow the tiny, 0.1\%, changes exhibited by
the Sun was a remarkable achievement.
In the meantime missions such as COROT \citep[COnvection, ROtation and
planetary Transits;][]{baglin-et-al-2002} and Kepler
\citep{borucki-et-al-2003} can detect similar levels of fluctuations on
myriads of stars, but reliably measuring the variability of the Sun remains
a particular challenge because of the immense brightness contrast of the Sun
compared with other astronomical objects, so that maintaining photometric
calibrations employing comparisons with many stars is difficult at best
(although some instruments do employ this technique in the UV).

\begin{figure}%
\epsfxsize25pc         %
\caption{
Space-borne total solar irradiance
(TSI) measurements covering the period 1978--2012 after the TRF (TSI
Radiometer Facility) corrections.
Individual records are shown in different colours, as labelled in the plot.
The bottom part of the plot shows the monthly mean sunspot number.
Courtesy of G.~Kopp (http://spot.colorado.edu/~koppg/TSI/).
}
\bigskip\bigskip
\centerline{\epsfbox{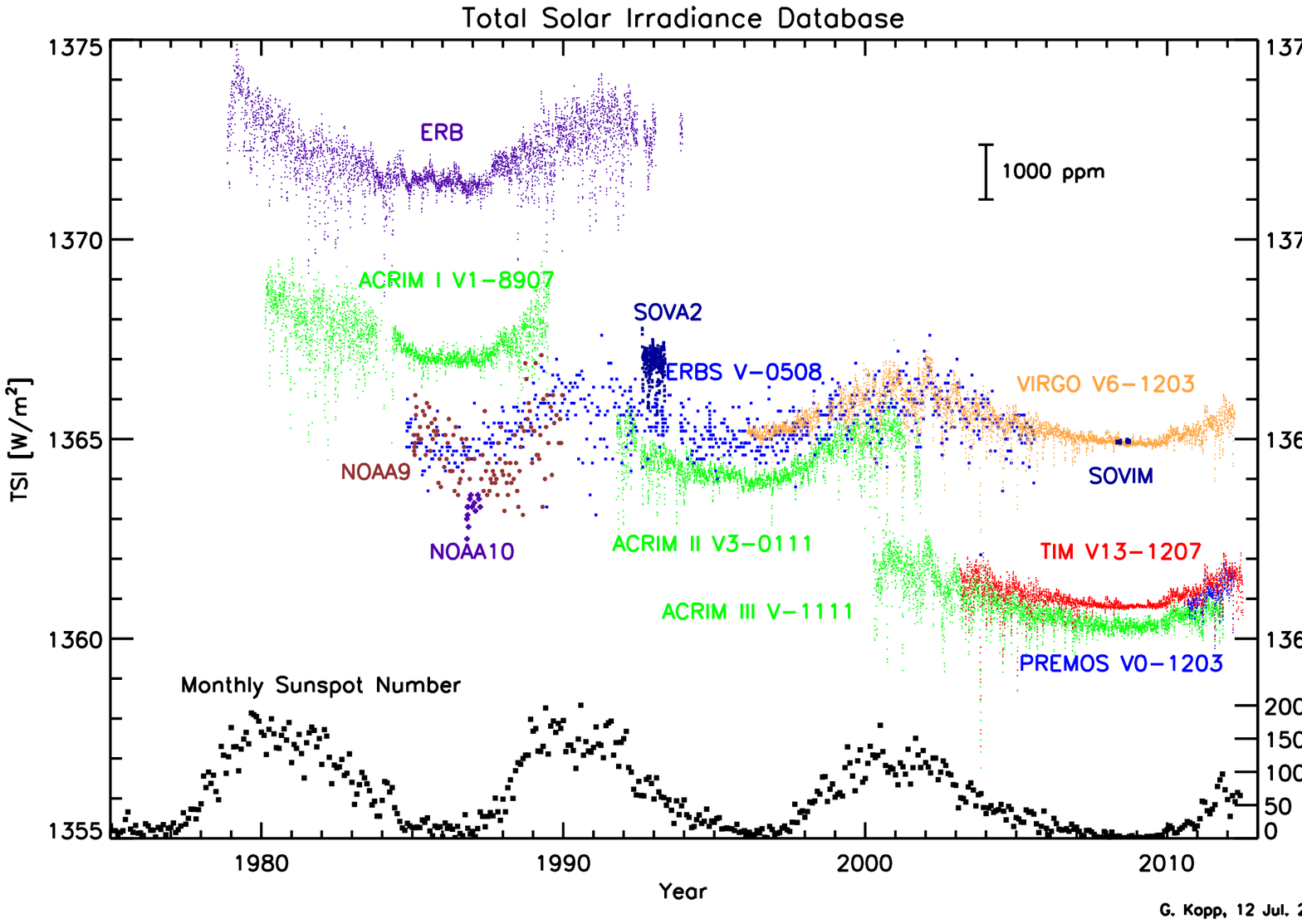}}
\label{tsi-kopp}
\end{figure}

Many attempts to measure ``the solar constant'' preceded the satellite
measurements that finally revealed its variability.  
Pre-satellite measurements of the solar constant obtained absolute values
ranging from 1338~Wm$^{-2}$ to 1428~Wm$^{-2}$ \citep[see reviews
by][]{smith-gottlieb-74,froehlich-brusa-81}.
These measurements were too inaccurate to detect intrinsic changes in the
solar brightness, even though their presence was suspected
\citep[e.g.,][]{eddy-76}.
Space-borne radiometers
\citep[e.g.,][]{hickey-et-al-80,willson-hudson-88,froehlich-2006}
provide an almost uninterrupted record of TSI since November 1978
(see Fig.~\ref{tsi-kopp}).
These instruments are accurate and stable enough to trace irradiance
{\it variability} up to the solar cycle timescale, as revealed in
Fig.~\ref{tsi-kopp} by the similarity of the curves recorded by different
instruments.
All the curves (with sufficient time
resolution) show two striking features: a trend following the
solar cycle, with the irradiance being higher during cycle phases with
higher activity and short (week-long) dips in irradiance that coincide with
the passage of sunspots across the solar disk. 

However, the radiometric accuracy of individual TSI measurements was
generally poorer than the $\sim0.1$\% solar cycle change, as indicated by the
scatter in absolute values in Fig.~\ref{tsi-kopp}.
A discrepancy of roughly 5~W/m$^2$, or 0.35\%, has been present between the
values measured by instruments launched in the 1980s and 1990s and the Total
Irradiance Monitor (TIM) on SOlar Radiation \& Climate Experiment (SORCE)
launched in 2003 (see Fig.~\ref{tsi-kopp}).
This discrepancy appears to have been resolved thanks to recent tests with
the TSI Radiometer Facility (TRF),
which allows TSI instruments to be validated
against a NIST-calibrated (National Institute of Standards and Technology)
cryogenic radiometer at full solar power under vacuum conditions before
launch.
It has been used to calibrate the PREcision MOnitor Sensor (PREMOS) on
PICARD (satellite named after the French astronomer Jean Picard), as well as
ground-based representatives of the Total Irradiance Monitor (TIM) on
SORCE, the Variability of Solar Irradiance and Gravity Oscillations
instrument (VIRGO) on Solar Heliospheric Observatory (SoHO) and the Active
Cavity Radiometer Irradiance Monitor (ACRIM) on ACRIM-sat
\citep{kopp-lean-2011,kopp-et-al-2012,fehlmann-et-al-2012}.
TRF-based tests and
corrections brought PREMOS and ACRIM3 into close agreement, to within
0.05\%, with TIM \citep{kopp-et-al-2012,fehlmann-et-al-2012}.
Thus a TSI value of 1360.8$\pm 0.5$~Wm$^{-2}$ is currently considered to
best represent solar minimum conditions \citep{kopp-et-al-2012}.

The discrepancy in absolute values of individual TSI measurements makes it
hard to assess irradiance changes on time
scales longer than the solar cycle, since individual radiometers rarely
covered more than a single solar activity minimum.
To trace long-term changes, individual irradiance measurements need to be
adjusted to the same absolute scale, which is a non-trivial task due to
instrumental degradation, sensitivity changes and other problems.
In particular, early sensitivity changes, when a radiometer starts being
exposed to sunlight need to be taken into account
\citep[e.g.,][]{lee-et-al-1995,dewitte-et-al-2004a,froehlich-2006}, but also
sudden changes in calibration or noise-level, like the ones that occurred on
the Nimbus-7 Earth Radiation Budget (ERB) instrument
\citep{hoyt-et-al-92,lee-et-al-1995}, can complicate a
``cross-calibration'' process.

\begin{figure}%
\caption{
Comparison of the three smoothed TSI composites (ACRIM, yellow; IRMB, green
and PMOD, red) as well as TSI reconstructed by the SATIRE-S (blue), all normalised to
SORCE/TIM at the minimum in December 2008.
Gaps in the curves are gaps in the data longer than 27~days.
Credit: \citet{ball-et-al-2012a},
reproduced with permission \copyright ESO.
}
\epsfxsize30pc         %
\centerline{\epsfbox{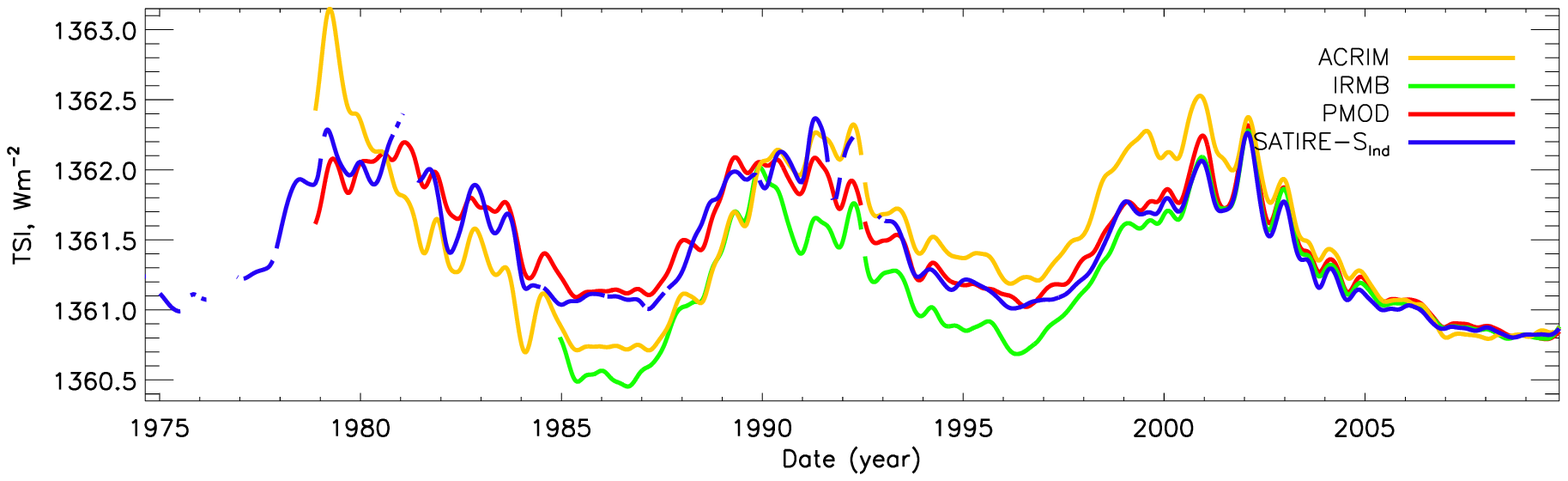}}
\label{fig-ball}
\end{figure}

Consequently, it is not surprising that
three different composites have been produced,
which are named after the instrument that they take as the basis, or the
institute at which the composite is produced: ACRIM
\citep{willson-97,willson-mordvinov-2003}, RMIB
 \citep[named after the Royal Meteorological Institute of Belgium, sometimes
also called IRMB in the francophonic tradition;][]{dewitte-et-al-2004} and PMOD
\citep[Physikalisch-Meteorologisches Observatorium Davos;][]{froehlich-2006}.
These composites are plotted in Fig.~\ref{fig-ball} after
imposing a temporal filtering to bring out the longer term changes.
The composites agree in many respects, e.g., on short time scales, and they
also share most of the features on longer time scales.

The most critical difference between them concerns their longer term trends,
which become most clearly visible by comparing the TSI levels during
activity minima, i.e.  at times when different levels
of TSI are easily distinguishable.
Such long-term changes are particularly interesting in the context of the
global climate change as witnessed in the last century,
which explains the debate that these differences in trend have sparked.
The ACRIM composite shows an upward trend between the minima in 1986 and 1996,
whereas TSI decreases from 1996 to 2008.
In the RMIB composite, TSI increases from the minimum in 1996 to 2008.
The PMOD TSI shows the opposite trend.
The differences in the composites and their sources are described in more
detail by \citet{froehlich-2006,froehlich-2012}.
Independent
models assuming irradiance variability to be driven by the evolution of the
surface magnetic field agree better with the PMOD long-term trend (see
Sect.~\ref{modelling}).

\begin{figure}%
\caption{Top: Reference solar spectrum recorded in April 2008
\citep{woods-et-al-2009}.
Bottom: Relative SSI variability as observed by UARS/SUSIM \citep[red
curve;][]{floyd-et-al-2003b}
between the maximum of cycle~23 (March 2000) and the
preceding minimum (May 1996),
as well as by SORCE/SOLSTICE \citep[light blue;][]{snow-et-al-2005}
and SORCE/SIM \citep[dark blue;][]{harder-et-al-2009}
between April 2004 and December 2008. 
Also shown is the variability between 2000 and 1996 predicted by the SATIRE model
\citep[green;][Sect.~\ref{modelling}]{krivova-et-al-2009a,krivova-et-al-2011a}.
For each period, averages over one month are used.
Negative values are indicated by dotted segments.
}
\epsfxsize25pc         %
\centerline{\epsfbox{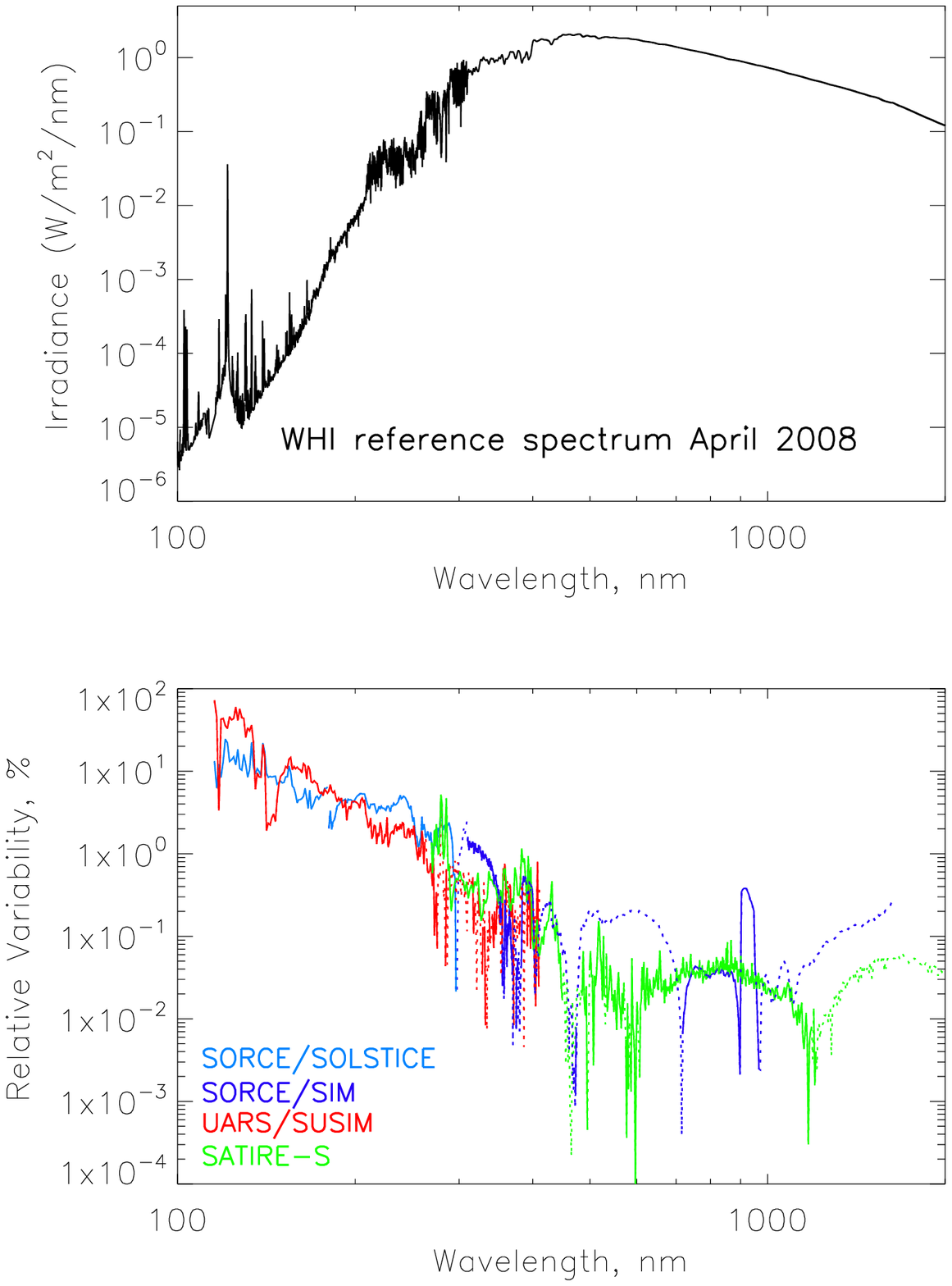}}
\label{fig-ssi-var}
\end{figure}

Space-based observations of SSI also started in 1978 with the Nimbus-7 Solar
Backscatter Ultraviolet radiometer \citep[SBUV;][]{cebula-et-al-92}, and
until 2002 were almost exclusively limited to the UV range below 400~nm.
They were reviewed by
\citet{deland-cebula-2008,deland-cebula-2012}.
\citet{deland-cebula-2008} have also collected the earlier
UV data and combined them into a single record.
The cross-calibration of individual data sets and construction of a
self-consistent composite is, however, in this case even more challenging
than for the TSI.

First observations in the visible and IR were sporadic \citep[e.g.,
by the SOLar SPECtrum, SOLSPEC and SOSP,
instruments flown on space shuttles or on the EUropean REtrieval CArrier
platform;][]{thuillier-et-al-2003,thuillier-et-al-2004a};
see also overviews by
\citet{thuillier-et-al-2009} and \citet{ermolli-et-al-2012a}.
These early measurements were used
to produce ATLAS (ATmospheric Laboratory for Applications and Science)
solar reference spectra ATLAS1 (March 1992) and ATLAS3 \citep[November
1994;][]{thuillier-et-al-2004a}.
More recent solar reference spectra were produced within the Whole
Heliosphere Interval (WHI) international campaign during three relatively
quiet periods in March--April 2008 \citep{woods-et-al-2009}.
One of these spectra, produced during the most quiet period
(the WHI quiet Sun reference spectrum), is shown in the upper panel of
Fig.~\ref{fig-ssi-var} (on a logarithmic scale to allow for the differences
in irradiance in the UV and the visible).  The strong emission line is
Ly~$\alpha$.

Assessment of the SSI variability is complicated and until relatively
recently, this was only possible in the UV range, mainly thanks to the
two instruments on board the Upper Atmosphere Research Satellite (UARS): the
Solar-Stellar Irradiance Comparison Experiment
\citep[SOLSTICE;][]{rottman-et-al-93} and the Solar Ultraviolet Spectral
Irradiance Monitor \citep[SUSIM;][]{brueckner-et-al-93}.
In the UV below about 250~nm, long-term instrumental uncertainties of
SOLSTICE and SUSIM were smaller than the solar variability
\citep[e.g.,][]{woods-et-al-96}.
The lower panel of Fig.~\ref{fig-ssi-var} illustrates the relative
difference in the irradiance spectrum between activity maximum and minimum
\footnote{Note that SORCE was launched in 2003 and SIM
data are only available since April 2004.
Thus SSI variability can be estimated from the SORCE data only over less
than half a cycle.}.
Clearly, the irradiance variability is a strong function of wavelength and
increases very rapidly towards shorter wavelengths in the UV
(note the logarithmic scale).
In spite of differences in detail, all data sets
show a qualitatively similar behaviour in the UV, illustrated by the red and
blue curves.
Quantitatively there are differences of up to nearly an order of magnitude,
which do not appear so striking due to the logarithmic scaling.
They are discussed below.

The results of the Solar Radiation and Climate Experiment (SORCE) launched
in 2003 sprang a surprise.
SORCE carries two instruments, the SOLSTICE \citep[an analogue of
UARS/SOLSTICE,][]{snow-et-al-2005} and the Spectral Irradiance Monitor
\citep[SIM,][]{harder-et-al-2005a}, which observe SSI over a broad spectral
range from Ly-$\alpha$ to 2400~nm.
Between 2003 and 2008, i.e.
over the declining phase of cycle~23, SIM displayed an anticyclic behaviour
in the visible, i.e. the irradiance at most visible and IR wavelengths is
lower at higher activity levels than during quiet times
\citep{harder-et-al-2009}. 
This is indicated by the dotted blue line in Fig.~~\ref{fig-ssi-var}.
These values whould be negative (which cannot
be directly represented on a logarithmic scale).
This anti-phase behaviour to the TSI is largely compensated by the enhanced
in-phase UV variability between 200 and 400~nm compared with previous
measurements.
Thus the estimated contribution of the 200--400~nm spectral range to the TSI
decrease from 2004 to 2008 was about 180\% \citep{harder-et-al-2009},
compared with 20-60\% based on earlier measurements
\citep{lean-et-al-97,thuillier-et-al-2004a,krivova-et-al-2006a,morrill-et-al-2011}.

At the same time, the short-term (rotational) variability measured by
SORCE/SOLSTICE and SORCE/SIM agrees with previous results very well.
Since shorter time scales are significantly less affected by instrumental
effects, \citet{deland-cebula-2012} conclude that undercorrection of
response changes for the SORCE instruments is the most probable source of
the discrepancies.

The importance of getting the correct spectral dependence of the irradiance
variations lies in the fact that the UV irradiance influences atmospheric
chemistry more strongly than that in the visible, although the visible
cannot be neglected (see Sect.~\ref{climate}).

\subsection{Physical Causes of Irradiance Variations}
\label{mecha}

\begin{figure}
\epsfxsize25pc         %
\caption{Power spectrum of total solar irradiance. The different
curves have the following meaning: Fourier (grey line) and global wavelet
(black dotted line) power spectra (in ppm$^2 /60 \mu$Hz) of the VIRGO data set
for the year 2002 sampled at a 1 min cadence (Fr\"{o}hlich et al. 1995). Black
solid line shows the
global wavelet spectrum of the SORCE TIM data \citep{kopp-et-al-2005b}
for the year 2003 sampled every 6~h.
Credit: \citet{Seleznyov-ea2011},  
reproduced with permission \copyright ESO.
}
\bigskip\bigskip
\centerline{\epsfbox{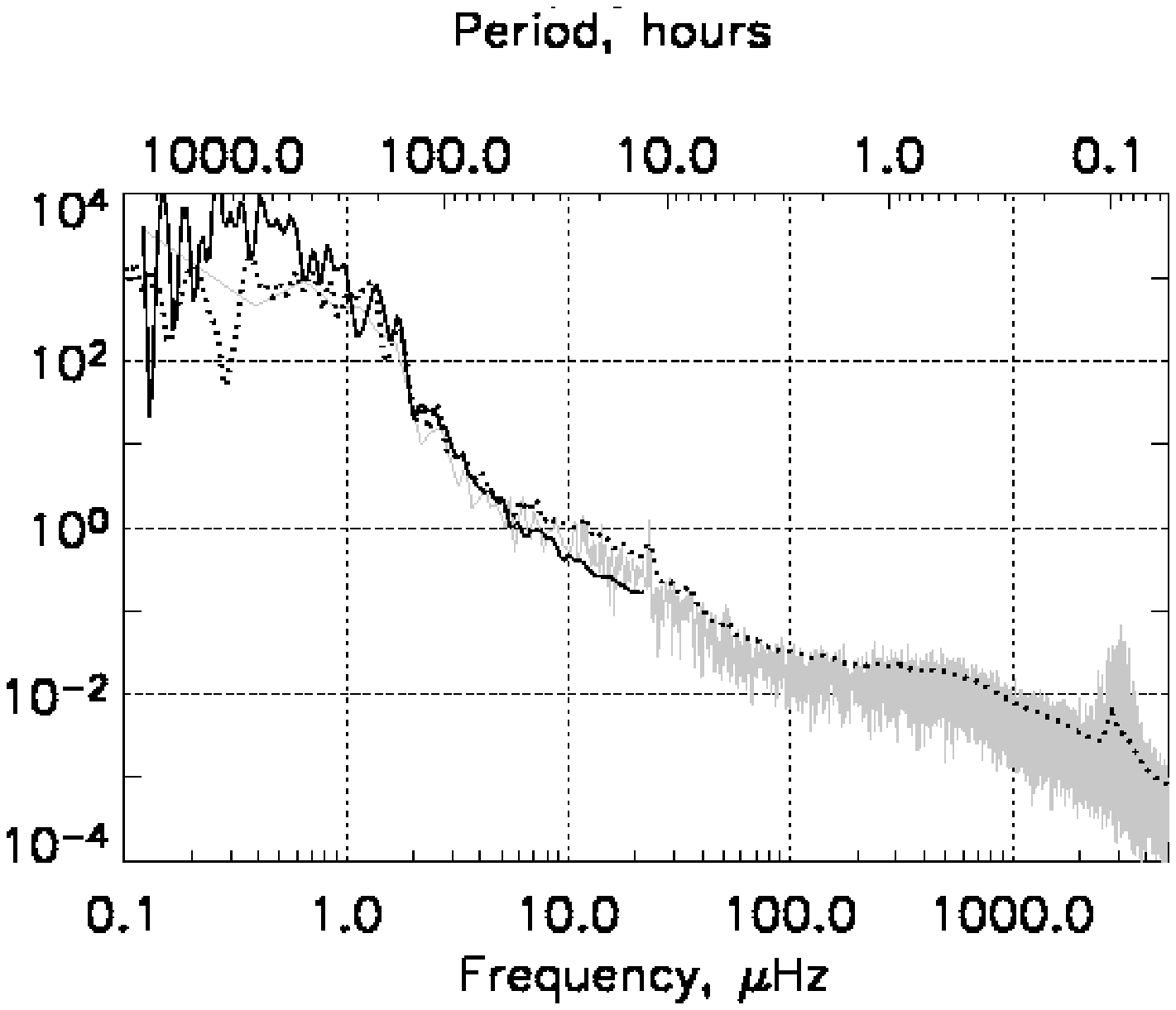}}
\label{Seleznyovfig}
\end{figure}

There are a variety of causes of solar irradiance variations, each acting on
particular timescales.
This is illustrated in Fig.~\ref{Seleznyovfig} for a limited set of
timescales by plotting the power spectrum of TSI for periods from about 1
minute to 1 year \citep{Seleznyov-ea2011}.
The power
drops very roughly as a power law from long to short time scales. These
variations are driven by a range of sources. Thus, solar oscillations, the
$p$-modes, are responsible for
the group of peaks centered on 5 min (3 mHz), the evolution of granules
produces the plateau between 50 and 500 $\mu$Hz (i.e. on periods of minutes to
hours), while the rotational modulation of TSI by the passage of sunspots and
faculae over the solar disc leads to the increase in power between 0.4 and
50 $\mu$Hz (i.e. periods of between about 5 and 5000 hours).
The solar rotation period itself does not display a significant peak in this
figure, however, due to the combination of the limited lifetime of sunspots
(most live only a few days or less) and the constant evolution of the
brightness of the longer lived active regions, as well as the rather common
occurance of multiple active regions on the Sun at the same time.

From the solar rotation period to the 10--12 year solar-cycle period the
growth, evolution and decay of active regions, as well as the distribution of
their remnant magnetic field over the solar surface provide the main contribution
to the variability. This is strongly modulated by the activity cycle itself,
which is a very strong contributer to irradiance variations.
Beyond the solar cycle period, the difference in the strength of individual
cycles as well as possible evolution of the background field and other
mechanisms (see Sect.~\ref{long-term}) may lead to a secular change, which
might be visible in the form of different TSI levels at the different minima.
Over the thermal relaxation time-scale of the convection zone of $10^5$ years
the energy blocked by sunspots should be gradually released again (see below),
while beyond $10^6$ years the gradual brightening of the Sun due to the chemical
evolution of its core should start to become noticeable in its TSI
\citep[e.g.,][]{Sackmann-ea1993,Charbonnel-ea1999,Mowlavi-ea2012}.

With the exception of the shortest and the longest time scales, the causes
of irradiance variations listed above are associated more or less directly
with the Sun's magnetic field mainly via the influence of magnetic fields on
the thermal structure of the solar surface and atmosphere.

For irradiance variations of possible relevance to global climate change,
the magnetic field is
expected to be the main driving force. Of importance is the magnetic field at
the solar surface and in the lower solar atmosphere, mainly the photosphere
\citep[see, e.g.,][]{solanki-unruh-98}. In these layers the magnetic field is
thought to be concentrated into strong field features \citep[having an average
field strength of roughly a kG in the mid photosphere;][]{Solanki-ea1999}
whose simplest description is by magnetic flux tubes (see
\citealt{solanki-93} for a
review), although their real structure is more complicated
\citep[e.g.,][]{Voegler-ea2005,Rempel-ea2009,Stein2012}. Another,
more chaotic component of the magnetic field is present as well (see
\citealt{deWijn-ea2009} for a review). It is still unclear if this turbulent
field component really contributes to irradiance variations, so that in the
following we restrict ourselves to the concentrated fields which range in
cross-section
size between structures well below 100 km in diameter to sunspots that often
have dimensions of multiple 10 Mm.

Sunspots, forming the hearts of active regions, clearly are dark (see
\citealt{solanki-2003,Rempel-Schliche2011} for reviews), while the
small magnetic elements that populate
(and form) the faculae in active regions and the network elsewhere on the Sun
\citep[and are even found in the internetwork of the quiet
Sun;][]{SanchezAlmeida-ea2004,Lagg-ea2010} are bright, particularly
near the limb and
at wavelengths formed above the solar surface.
Such wavelengths include the Fraunhofer g-band
\citep{Muller-Roudier1984,Berger-ea1995}, the CN bandhead
\citep{Sheeley1969,Zakharov-ea2007}, the cores of strong spectral
lines \citep[e.g.,][]{Skumanich-ea1975} and the UV
\citep{riethmueller-et-al-2010}.

The darkness of sunspots is due to the blocking of heat flowing from below by the
kG magnetic field, which is strong enough to largely quench overturning
convection \citep{Rempel-Schliche2011}.
Forms of magnetoconvection do take place, maintaining the penumbra's (and to
some extent also the umbra's) brightness
\citep[e.g.,][]{schuessler-voegler-2006,Rempel-ea2009,%
Joshi-ea2011,Scharmer-ea2011}.

\begin{figure}
\epsfxsize20pc         %
\centerline{\epsfbox{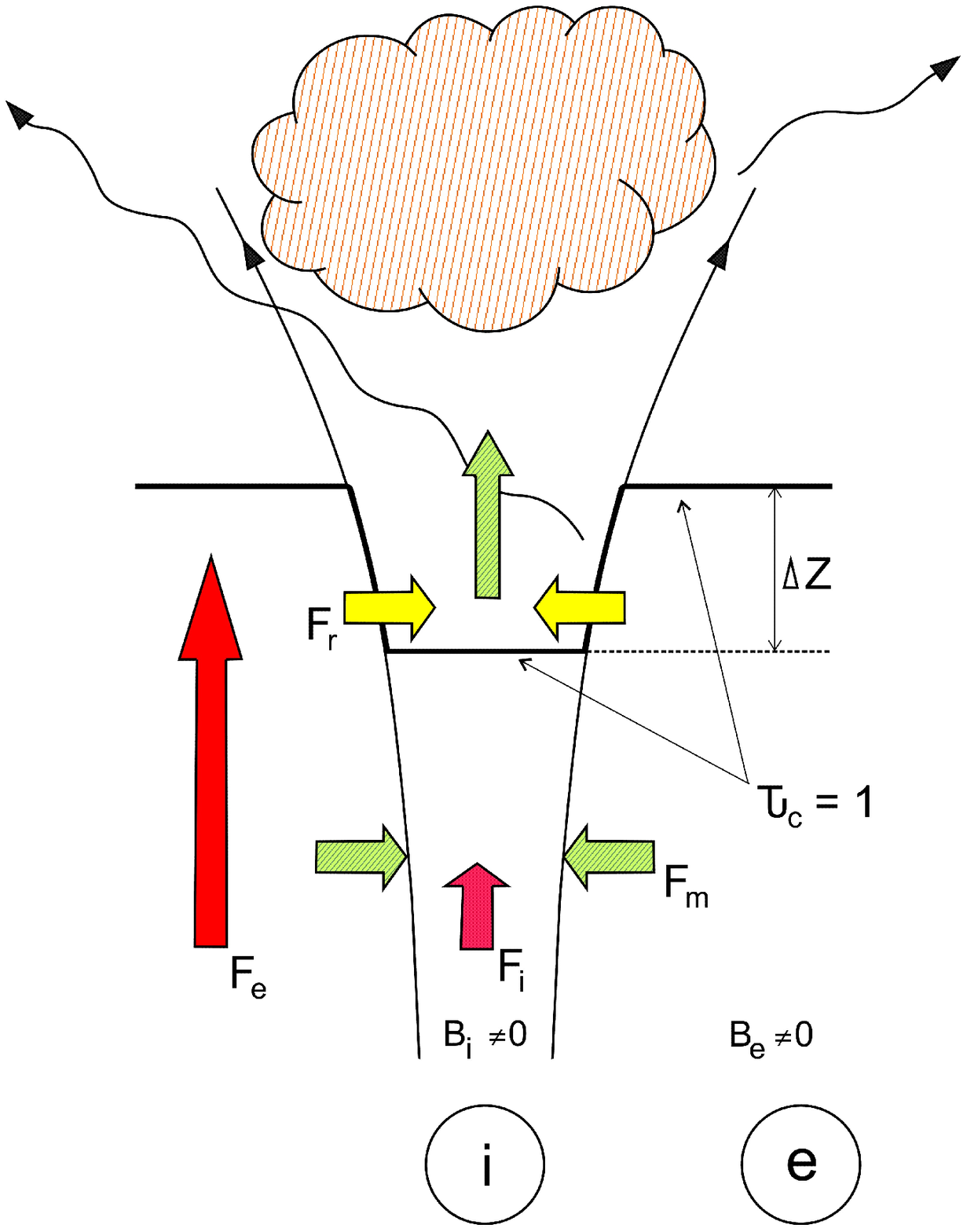}}
\caption{Sketch of the vertical cross-section through a slender magnetic flux
tube. The arrows illustrate the various forms of energy
transfer. Red arrows: vertical convective and radiative energy flux below
the solar surface inside
the flux tube (subscript i) and in the external medium (subscript e). Yellow
arrows: horizontal influx of radiation through the walls
of the flux tube (the thick lines outline the optical depth unity, $\tau =1$,
surface, as seen from above). $\Delta Z$ represents the Wilson depression.
Green arrows: mechanical energy flux. The cloud sketches the hot
chromospheric layers of the magnetic feature (roughly following a sketch by
\citealt{zwaan-78}).}
\label{fluxtube}
\end{figure}

The magnetic elements have a nearly equally strong field (as sunspots
averaged over their cross-section), so that the convective energy flux from
below is greatly reduced in their interiors. This is indicated by the
different lengths of the vertical red arrows in Fig.~\ref{fluxtube}. This
figure displays the vertical cross-section of an intense slender flux tube.
Another feature that can be seen in the figure is the depression of the
optical depth unity surface (heavy black line) in the flux tube's interior.
Hence, for depths up to $\Delta Z$ below the solar surface in the quiet Sun,
the walls of the flux tube allow radiation to escape into space. This
radiation heats up the interior of the tube.
Forthermore, photons from these hot walls (which are windows
into the hot interior of the Sun) can be directly observed, best
when the magnetic element is located at some distance away from solar disc
centre \citep{Spruit1976,Keller-ea2004,Carlsson-ea2004}.

In addition to this radiative heating, the magnetic elements are shaken and
squeezed by the turbulently convective gas in their surroundings. This causes
the excitation of different wave modes within them
\citep{musielak-ulmschneider-2003}.
This mechanical transfer of
energy from the surroundings into the tubes is represented in
Fig.~\ref{fluxtube} by the
green horizontal arrows, while the upward transport of the mechanical energy
by waves is indicated by the vertical green arrow.

In particular the longitudinal tube waves steepen as they propagate upwards,
due to the drop in density, finally dissipating their energy at shocks in
the chromosphere \citep[e.g.,][]{Carlsson-Stein1997,Fawzy-ea2012}. Other
forms of heating may also be taking place, but are not discussed further
here. This leads to a heating of the upper photospheric and chromospheric
layers of magnetic elements, which explains their excess brightness in the
UV and in the cores of spectral lines \citep[e.g. Ca~II H and K,
see][]{Schrijver-ea1989,Rezaei-ea2007}.

Whereas the radiation flowing in from the walls penetrates the small magnetic
features completely, for features with horizontal dimension greater than
roughly 400 km the radiation cannot warm the inner parts and they remain cool
and \nocite{grossmann-doerth-ea1994} dark, cf.~Grossmann-Doerth
et al.~(1994).

The radiative properties of sunspots (and to a lesser extent the smaller,
but still dark pores) and magnetic elements are responsible for most of the
irradiance variations on time scales of days to the solar cycle and very
likely also beyond that to centuries and millenia, as
described in the following sections.

However, one important question remains: Why does the energy blocked by the
magnetic field in sunspots not simply flow around them and appear as a
surrounding bright ring? Such bright rings have been
found \citep[e.g.,][]{Waldmeier1939,Rast-ea2001}, but prove to release only
a few percent of the energy flux blocked by the enclosed sunspots. An
explanation was provided by \citet{Spruit1982a,Spruit1982b}, who showed that
the energy blocked by sunspots is redistributed in the solar convection zone
due to the very high heat conductivity of the solar plasma (see
\citealt{Spruit2000} for a review). Due to the very high heat capacity, the
stored heat hardly changes the surface properties at all. This heat is
gradually re-emitted over the Kelvin-Helmholtz timescale of the convection
zone, which is approximately $10^5$~yr. In analogy, the excess radiation
emitted by magnetic elements also comes from the convection zone's large
heat reservoir. Consequently, one may consider magnetic elements as leaks in
the solar surface, since by dint of being evacuated they increase the solar
surface area (see Fig.~\ref{fluxtube}).

Alternative explanations to the TSI variations on timescales of
the solar cycle and longer have also been proposed, e.g. thermal
shadows produced by horizontal magnetic flux in the solar interior
\citep{Kuhn-ea1998} 
or long-period ($\approx 1$ month) oscillations driven by the Coriolis force
\citep[r-modes;][]{Wolff-Hickey1987}.
As we shall see in the next subsection, the surface magnetic field leaves at
most a few percent of the TSI variations on timescales up to the solar cycle
to be explained by these or other such mechanisms.
However, it cannot be ruled out that one of these mechanisms
(or an as yet unknown one) may contribute
significantly on longer time scales \citep[e.g.,][]{Sofia-Li2001}.


\subsection{Modelling of TSI and SSI}
\label{modelling}

Models assuming that irradiance variations on time scales longer than
roughly a day are caused by changes in the surface distribution of different
magnetic features (see Sect.~\ref{mecha}) turned out to be most successful in
explaining observed irradiance changes.
The first models of this type
\citep[e.g.,][]{willson-et-al-81,oster-et-al-82,foukal-lean-86} were
so-called proxy models, which combined proxies of solar surface magnetic
features using regressions to match observed TSI changes.
Proxies that have been used most frequently include the sunspot area and the
Photometric Sunspot Index (PSI) derived from it (a measure of sunspot
darkening), as well as Mg~II, Ca~II and F10.7 (solar radio flux at 10.7~cm)
indices to describe facular brightening.
The widely used Mg II index is the ratio of
the brightness in the cores of the Mg II lines to their wings, making it
relatively insensitive to instrumental degradation with time
\citep{viereck-et-al-2001}.
However, it is sensitive to the exact wavelength choice, which leads to
uncertainties in the long-term trend in the composite Mg~II record.
The Ca II index is similarly defined.
More accurate proxy models employ spatially resolved observations of the
full solar disc, which account for the center-to-limb variation
of spot and facular contrasts at least at one wavelength
\citep[e.g.][]{chapman-et-al-96,chapman-et-al-2012,preminger-et-al-2002}.
More details on proxy models can be found in reviews by
\citet{froehlich-lean-2004} and \citet{domingo-et-al-2009}.

With time, more physics-based models have been developed
\citep[e.g.,][]{fligge-et-al-2000a,krivova-et-al-2003a,%
wenzler-et-al-2006a,shapiro-et-al-2010a,ermolli-et-al-2011,fontenla-et-al-2011}.
They still use different (spatially resolved or disc-integrated)
observations or proxies of solar magnetic activity to describe the evolution
of the surface coverage by different types of solar feautures (such as spots
or faculae), also called components of the solar atmosphere.
But the brightness of each component is calculated using radiative
transfer codes from semi-empirical
models of different features in 
the solar atmosphere \citep[see, e.g.,][]{kurucz-93,fontenla-et-al-99,%
unruh-et-al-99,fontenla-et-al-2009,shapiro-et-al-2010a}.
Brightnesses of the photospheric components computed in this way depend on
the wavelength and the heliocentric position.
This brings two advantages: 1) it allows calculating the spectral
irradiance, which is less straightforward with proxy models, and 2) it takes
into account the centre-to-limb variation of the contrasts of different
magnetic components, which provides more accurate reconstructions of solar
irradiance.
A successful example is  the SATIRE-S (Spectral And Total Irradiance
REconstructions for the Satellite era) model employing daily solar
magnetograms and continuum images
\citep{krivova-et-al-2003a,wenzler-et-al-2006a,ball-et-al-2012a}.

\begin{figure}%
\caption{Top: TSI between 1978 and 2009 as given by the PMOD composite
record of measurements (light red curve for daily and thick red for
smoothed data) and as computed with the SATIRE-S model (light blue and thick blue
for daily and smoothed data, respectively).
Both data sets are normalised to SORCE/TIM data in December 2008.
The thin blue lines indicate the uncertainty range of the model.
Dashed horizontal lines mark TSI levels at cycle minima.
Dotted vertical lines indicate cycle maxima and minima.
Black error bars are the PMOD TSI errors from \citet{froehlich-2009}.
Bottom: the difference between PMOD and SATIRE-S TSI (daily, grey; smoothed,
black). The PMOD TSI error bars are shown in red.
Credit: \citet{ball-et-al-2012a},  
reproduced with permission \copyright ESO.
}
\epsfxsize35pc         %
\centerline{\epsfbox{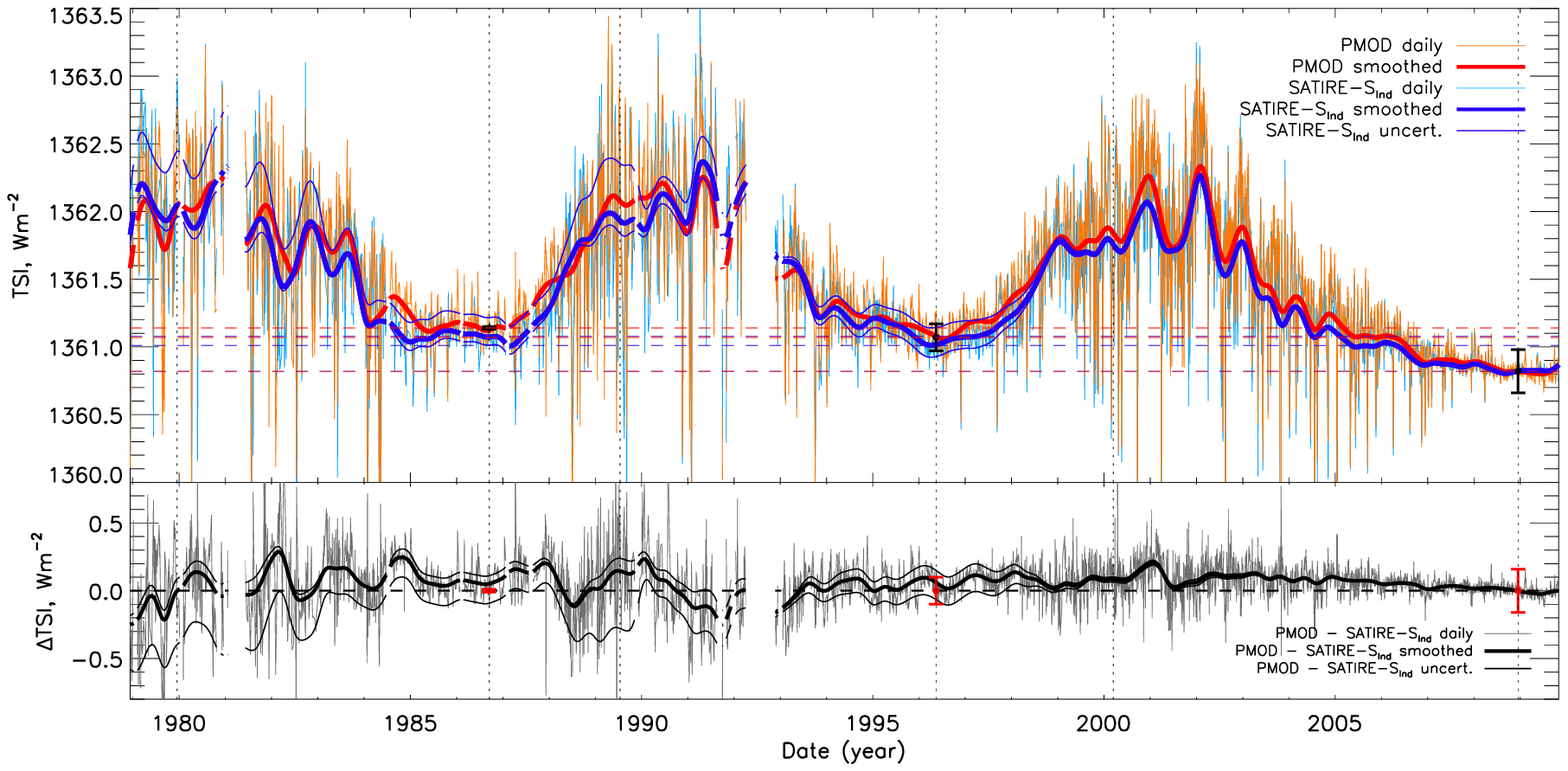}}
\label{fig-ball-pmod}
\end{figure}

In the last decade, significant progress has been made in modelling TSI
\citep{preminger-et-al-2002,ermolli-et-al-2003,krivova-et-al-2003a,%
lean-et-al-2005,wenzler-et-al-2006a,ball-et-al-2012a,chapman-et-al-2012}.
State-of-the-art models reproduce more than 90\% of the measured TSI
variations over the whole period covered by observations (see
Fig.~\ref{fig-ball-pmod}) and more than 95\% for cycle~23
\citep{ball-et-al-2011,ball-et-al-2012a,chapman-et-al-2012}, when compared
with the PMOD composite (see Sect.~\ref{obs}).

The models can in principle be used to distinguish between the composites.
On the solar cycle and longer time scales, \citet{wenzler-et-al-2009a},
\citet{krivova-et-al-2009b} and \citet{ball-et-al-2012a} found the SATIRE-S
model to be in best agreement with the PMOD composite (Fig.~\ref{fig-ball}),
although after the removal of long-term trends, the best agreement is
reached with the RMIB composite.
\citet{kopp-lean-2011} obtain similar correlations between their proxy
model (NRLSSI, see later in this section)
and the RMIB ($r_c$ = 0.92) and PMOD ($r_c$ = 0.91) composites.
%


Development of climate models including chemistry and thus increased
interest in solar UV data has stimulated advances to SSI modelling.
Over the years a number of models describing the variation of SSI have been
constructed.
These include
the Naval Research Laboratory Solar Spectral Irradiance model
\citep[NRLSSI;][]{lean-et-al-97,lean-2000}, SATIRE-S
\citep
{krivova-et-al-2006a,krivova-et-al-2009a,krivova-et-al-2011a,ball-2012}, the
COSI
\citep[COde for Solar Irradiance;][]{shapiro-et-al-2010a,shapiro-et-al-2011a},
SRPM  \citep[Solar Radiation Physical Modelling;][]{fontenla-et-al-99,%
fontenla-et-al-2004,fontenla-et-al-2009,fontenla-et-al-2011} and OAR
\citep[Osservatorio Astronomico di
Roma;][]{ermolli-et-al-2011,ermolli-et-al-2012a} models.
Two recent models by \citet{bolduc-et-al-2012} and
\citet{thuillier-et-al-2012} are limited to the UV spectral range only.
An overview of SSI models has recently been given by \citet{ermolli-et-al-2012a}.

These models reach different levels of complexity, make partly different
assumptions (with the common main underlying assumption being that evolution
of the magnetic field at and above the solar surface is the main cause of
SSI variability) and show partly significant differences in their results
(as discussed below).
Nonetheless, they do have some important traits in common that are worth
stressing (with one exception, the SRPM model by
\citealt{fontenla-et-al-2011}, discussed later in this section).


\begin{figure}%
\caption{
Normalised solar UV irradiance between 220 and 240~nm calculated
with NRLSSI (black), SATIRE-S (blue) and COSI (magenta), and measured
with UARS/SUSIM (darker green), UARS/SOLSTICE (light green), SORCE/SOLSTICE
(orange) and SORCE/SIM (red).
The pale green shading marks the period when the sensitivity of the
UARS/SUSIM instrument (and thus the flux) changed, so that a shift was
applied to the data before that (see
\citealt{krivova-et-al-2006a,krivova-et-al-2009a} for details).
Left-hand panel is limited to the period when SORCE was in operation, i.e.   
after 2003, and shows daily values, except for the COSI model, for which
only yearly averages are available.
Right-hand panel shows 3-month smoothed values over the period 1993--2009,
for which UARS and/or SORCE data are available.
From \citet{ermolli-et-al-2012a}.}
\centerline{\psfig{figure=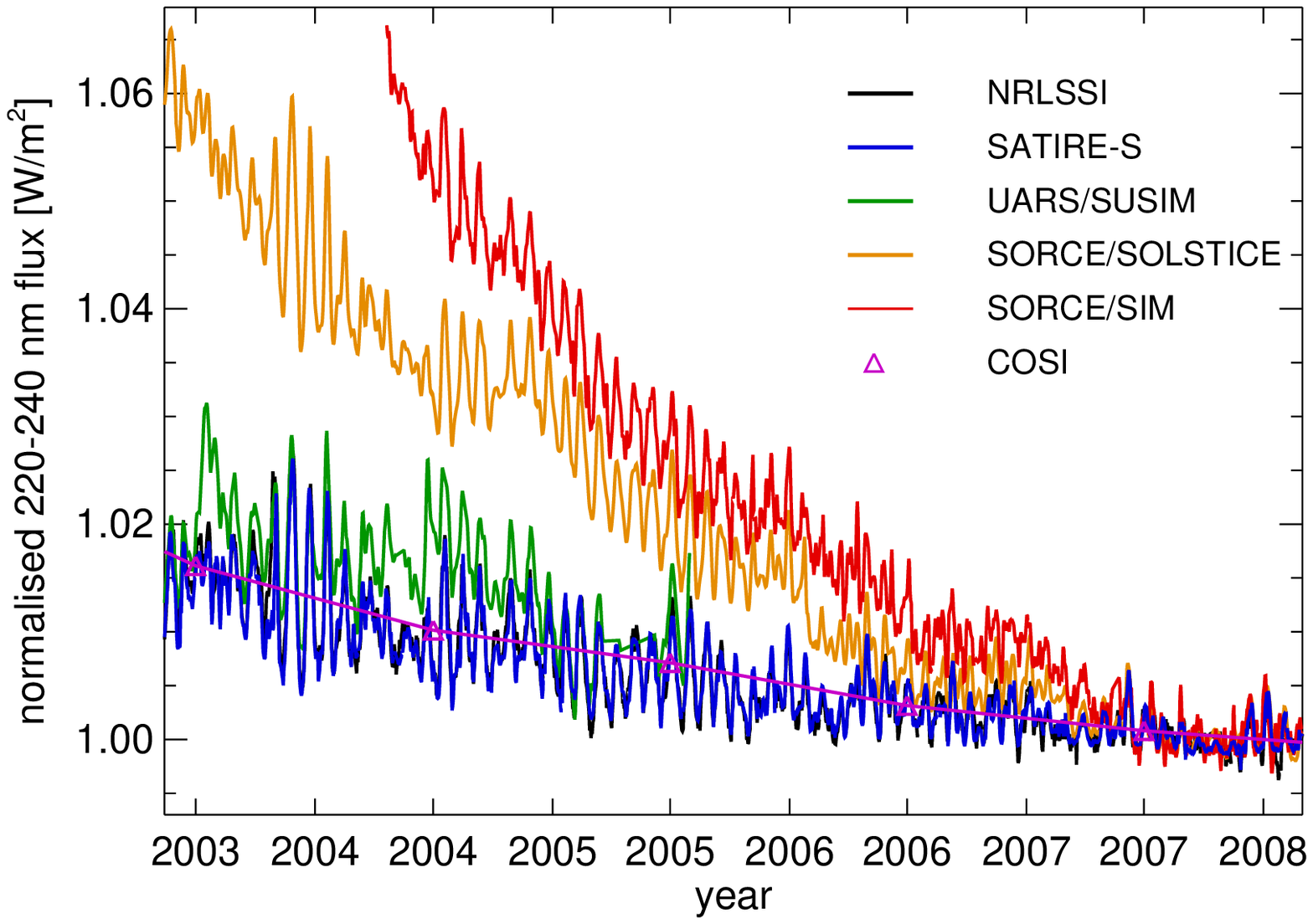,width=18pc}
\psfig{figure=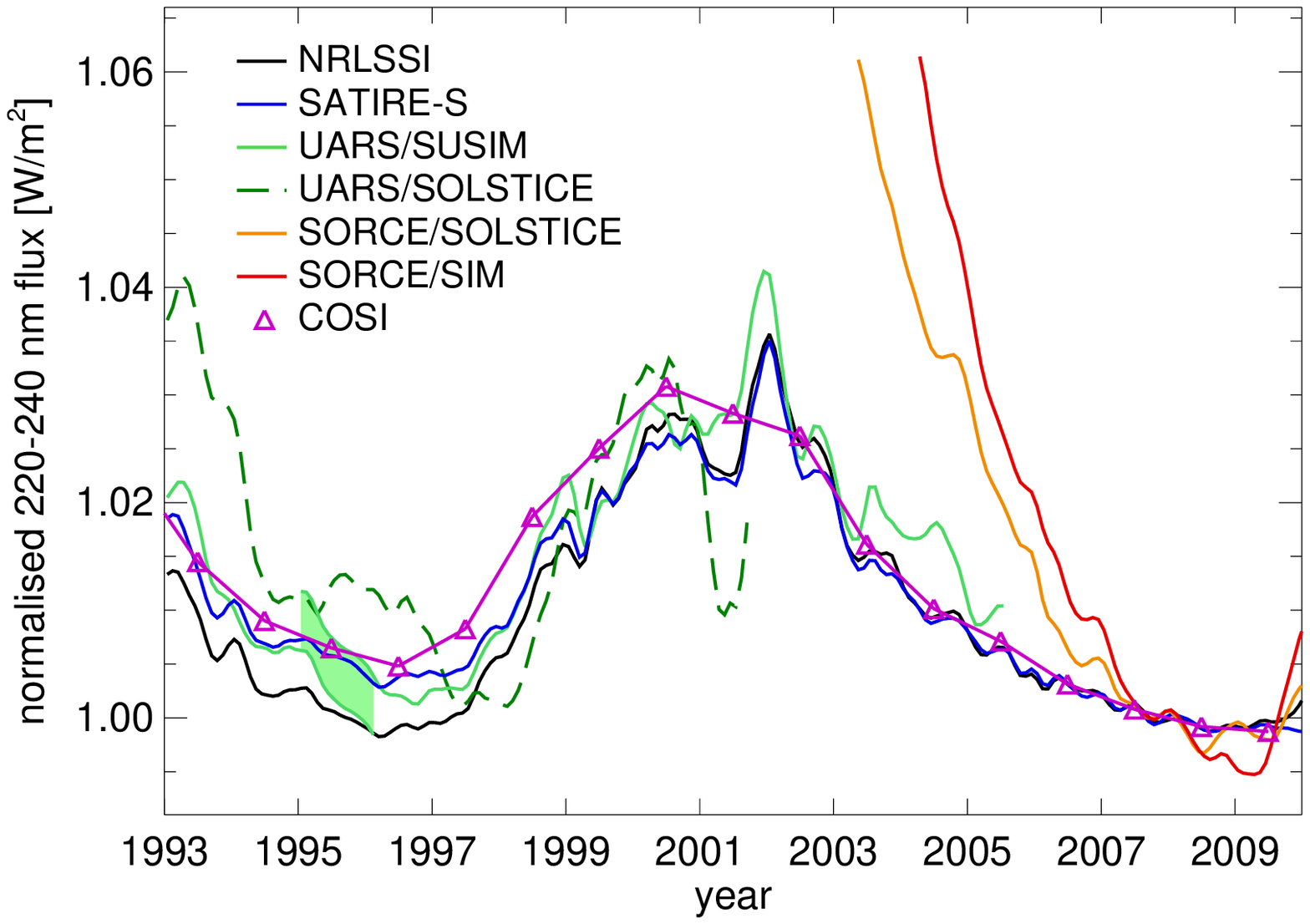,width=18pc}
}
\label{fig-yvonne}
\end{figure}

\noindent
1. All models produce a generally increasing level of SSI variability with
decreasing wavelength in the UV.  This is in qualitative agreement with the
measurements (see bottom panel of Fig.~\ref{fig-ssi-var}), while
quantitative comparisons are discussed below.

\noindent
2. On solar rotation timescales the models and data agree remarkably
well, at least for the more  advanced models.
This is illustrated by the left-hand panel of Fig.~\ref{fig-yvonne}, which
shows daily normalised
SORCE data (SIM in red and SOLSTICE in orange), UARS/SUSIM data
(green) and four different models in the spectral range
220--240~nm over the period 2003--2009.

\noindent
3. On the solar cycle time-scale the UV variability displayed by all
models is much lower (by factors of 2--6) than that shown by the SSI
instruments on SORCE, although for the more advanced models it agrees rather
well with the variations found by the SSI instruments on UARS.
The right-hand panel of Fig.~\ref{fig-yvonne} shows 3-month smoothed values
of solar irradiance at 220--240~nm between 1993 and 2009, calculated with
the NRLSSI (black curve), SATIRE-S (blue) and COSI (magenta; yearly values)
models and measured by UARS/SOLSTICE (darker green), UARS/SUSIM (lighter
green), SORCE/SOLSTICE (orange) and SORCE/SIM (red).
In this spectral range, all three models agree well with
each other over the whole period 1993--2008, as well as with the UARS data
between 1993 and 2005 (SUSIM stopped its operation in August 2005,
UARS/SOLSTICE in 2002).
But the 220--240~nm flux measured by the SORCE instruments over the period
2004--2008 decreased by a factor of 4 (SOLSTICE) to 7 (SIM) more than
expected from the models.
The difference between the trend measured by SORCE/SOLSTICE and
reconstructed by the models
actually lies within the 3-$\sigma$ long-term instrumental uncertaity
\citep{unruh-et-al-2012}. The discrepancy between the models and SIM is
larger, but in this spectral range SIM is considered to be less accurate
than SOLSTICE.

\noindent
4. The SSI variability is in phase with the solar cycle at all
wavelengths (with the exception of a short stretch in the IR).
This disagrees with the data from SORCE/SIM (cf. the green curve, showing
the SATIRE-S model, and the dotted part of the blue curve showing SIM data
at antiphase with the cycle in Fig.~\ref{fig-ssi-var}).


Although qualitatively the results  of most models are similar,
they also show significant quantitative differences.
Most important for climate models
is the discrepancy in the estimated UV variability at
250--400~nm, where models differ by up to a factor of three, e.g.  between
NRLSSI and COSI.
On the one hand, proxy models
generally tend to underestimate the variability in this range.
Such models rely on SSI measurements, at least in this spectral range
(e.g., NRLSSI and the models by
\citealt{pagaran-et-al-2009} or \citealt{thuillier-et-al-2012}),
and extrapolate observed rotational variability to longer time scales.

On the other hand, model atmospheres of the solar features employed in
semi-empirical SSI models (e.g., in SATIRE-S, COSI, SRPM, OAR) have not yet
been tested observationally at all wavelengths (due to the lack of
appropriate observations) and have some freedom as well, although they
cannot be tuned arbitrarily.
An example is given by the SRPM model \citep{fontenla-et-al-2011}, in
which the atmospheric models were tuned to allow better
agreement with the SORCE SSI data.
Thus it is the only model, which qualitatively reproduces SORCE/SIM SSI
behaviour, in particular, the reversed variability in the visible.
This is, however, achieved at the expense of TSI: the modelled TSI does not
reproduce the solar cycle change, which is measured much more reliably
than the SSI.
Interestingly, the OAR model uses essentially the same input as SRPM, with
the earlier untuned versions of the same atmospheric models
\citep{fontenla-et-al-2009} and is able to reproduce the TSI changes,
but not the SORCE SSI variability \citep{ermolli-et-al-2012a}.
The difference in the predicted variability at 250--400~nm between
semi-empirical models (excluding SRPM, i.e.  considering only models that
reproduce the TSI variability) is still almost a factor of two, with COSI
showing the strongest variability and OAR the weakest.  A more detailed
comparsion of the models has been presented by \citet{ermolli-et-al-2012a}.



\section{LONGER TERM SOLAR VARIABILITY: SECULAR CHANGE OF IRRADIANCE}
\label{long-term}

\subsection{Grand Maxima and Minima}
\label{grand}

\begin{figure}
\epsfxsize35pc         %
\centerline{\epsfbox{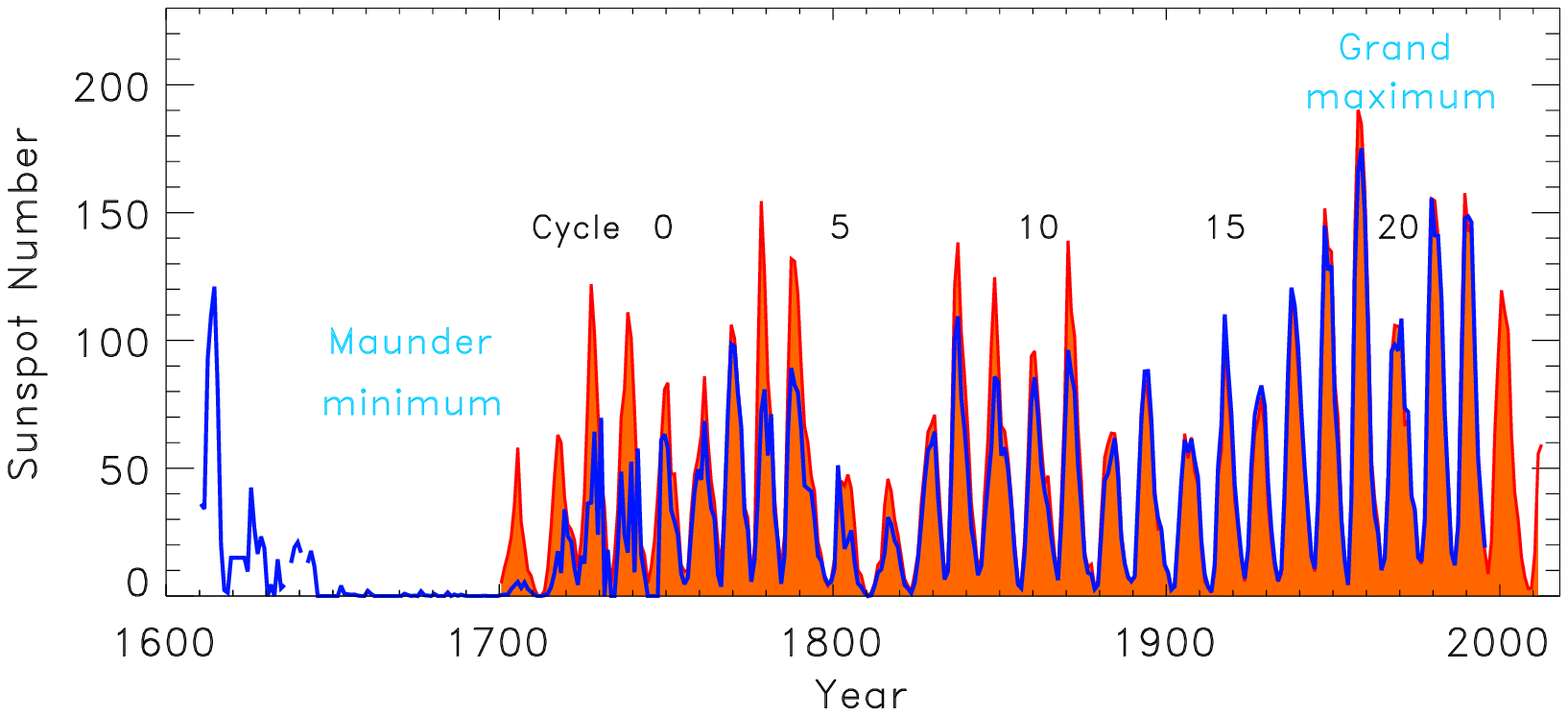}}
\caption{Telescopic, yearly averaged, sunspot number records: Zurich (orange)
and group sunspot number
\citep[blue;][]{vaquero-et-al-2011,hoyt-schatten-98,sidc-2011}.
The Zurich (or Wolf) number was introduced by Rudolf Wolf in the 1840s, the
group number by \citet{hoyt-schatten-98}.
The group sunspot number has been proposed
to better represent actual level of activity
before 1880, but is not yet officially available for cycles~23 and 24.}
\label{sunspotnumber}
\end{figure}

The longest running record of solar activity, available
since 1610, is the sunspot number (observations started only one year after the
invention of the telescope in 1609). It is a simple measure of the Sun's
activity, but nonetheless rather robust, especially when only sunspot groups are
counted \citep[as is the case for the group sunspot number introduced
by][]{hoyt-schatten-98}.
Robustness is a necessary condition since data from
different sources need to be combined in building up any long-running solar
activity record. The most striking feature of this record, plotted in
Fig.~\ref{sunspotnumber},
is the solar activity cycle. Each cycle lasts between 8 and 14 years, with an
average length of approximately 11.2 years.
Cycle amplitudes vary even more strongly, with the weakest known cycle
(starting around 1700)
being less than 10\% in strength of the strongest cycle, cycle 19, although
in the minima between the cycles the sunspot number reaches nearly zero
(there are some differences between minima following very strong
and those following weaker cycles).
The solar cycle has been reviewed in detail by \citet{hathaway-2010}.

Almost as striking as the presence of the cycles is their absence, along with
the near absence of sunspots themselves, between roughly 1640 and 1700
\citep{Eddy1976,Soon-Yaskell2003}.
This Maunder minimum is a prime
example of a grand minimum of solar activity, which in this case overlapped with a
particularly cold part of the little ice age. Grand minima of solar activity
have been reviewed by \citet{Usoskin2012}.

The only direct records of solar activity measurements that reach
further back in time are scattered naked-eye sightings of sunspots, mainly
in China \citep{Yau-Stephenson1988}.  These are too
scarce to allow a reliable reconstruction of solar activity.
An alternative is provided by cosmogenic isotopes, such as
$^{14}$C and $^{10}$Be stored in terrestrial archives. These isotopes are
produced in the Earth's atmosphere by nuclear reactions (neutron capture,
spallation) between energetic cosmic rays and
constitutents of the Earth's atmosphere (mainly N, but also O and Ar in the
case of $^{10}$Be).
After production the two main cosmogenic isotopes take different paths.
$^{14}$C becomes part of the global
carbon cycle until it ends up in one of the sinks, e.g. the ocean or in plant
material. Of interest are those atoms that end up in the trunks of datable
trees. After circulating in the atmosphere for a few years,
if it was formed in the stratosphere, $^{10}$Be precipitates and can be recovered (and
dated) if it is deposited on striated ice sheets, such as those of
Greenland or Antarctica.

 Since both isotopes are radioactive
(half-lives of 5730 years for $^{14}$C, and $1.36\times 10^6$ years for $^{10}$Be)
and have no terrestrial sources, their concentration in a layer
(or year-ring) of their natural archives is a measure of the production rate of
these isotopes at the time of deposition (after relevant corrections). This in turn
depends on the flux of cosmic rays reaching the Earth's atmosphere, which is
mainly determined by the strength (and, at least for $^{10}$Be, the geometry) of the
Earth's magnetic field, and on the level of solar activity. Hence, if the
geomagnetic field is known from a previous reconstruction
\citep[e.g.][]{Korte-Constable2005,Knudsen-ea2008},
then the level of solar activity, primarily modulation potential, which
depends mainly on the
Sun's open magnetic flux, can be determined using a simple model.

A further important requirement for $^{14}$C is that ocean circulation remains
roughly unchanged \citep[which can be shown for the Holocene,][]{stuiver-91}.
The $^{10}$Be, in turn, may be sensitive to variations in local climate
\citep[e.g.][]{Field-ea2006}.

The connection of the open magnetic flux with the quantities needed to
estimate solar irradiance variations, sunspot and facular areas, or
alternatively sunspot number and total magnetic flux, can then be
established via another simple model introduced by
\citet{solanki-et-al-2000,solanki-et-al-2002}, cf. 
\citet{vieira-solanki-2010}.
The main ingredient of this model is that solar cycles overlap due to two
mechanisms: (1) magnetic flux belonging to the new cycle starts to emerge
while flux belonging to the old cycle is still emerging and (2) magnetic
flux organized on large scales (i.e.  mainly the open flux) decays very
slowly, over a timescale of years, and hence is still present when the next
cycle is well underway.

\citet{Usoskin-ea2004} have shown that by putting together the
various models solar activity can be reconstructed reliably, at least from
$^{14}$C. This was put to use by \citet{Usoskin-ea2003}, who
reconstructed sunspot number for the last 1000 years, and by
\citet{solanki-et-al-2004},
who did that for the last 11400 years, i.e. basically the full
holocene. Reconstructions of solar activity based on $^{10}$Be followed later
\citep{steinhilber-et-al-2008,steinhilber-et-al-2009}.
Although the reconstructions based on the two
isotopes (and to some extent those based on different geomagnetic field
reconstructions) differ from each other in detail, they show a similar
statistical behaviour \citep{Usoskin-ea2009,steinhilber-et-al-2012}.
Such model-based reconstructions supercede earlier work by, e.g.,
\citet{Stuiver-Braziunas1989} with $^{14}$C.

\begin{figure}
\epsfxsize30pc         %
\centerline{\epsfbox{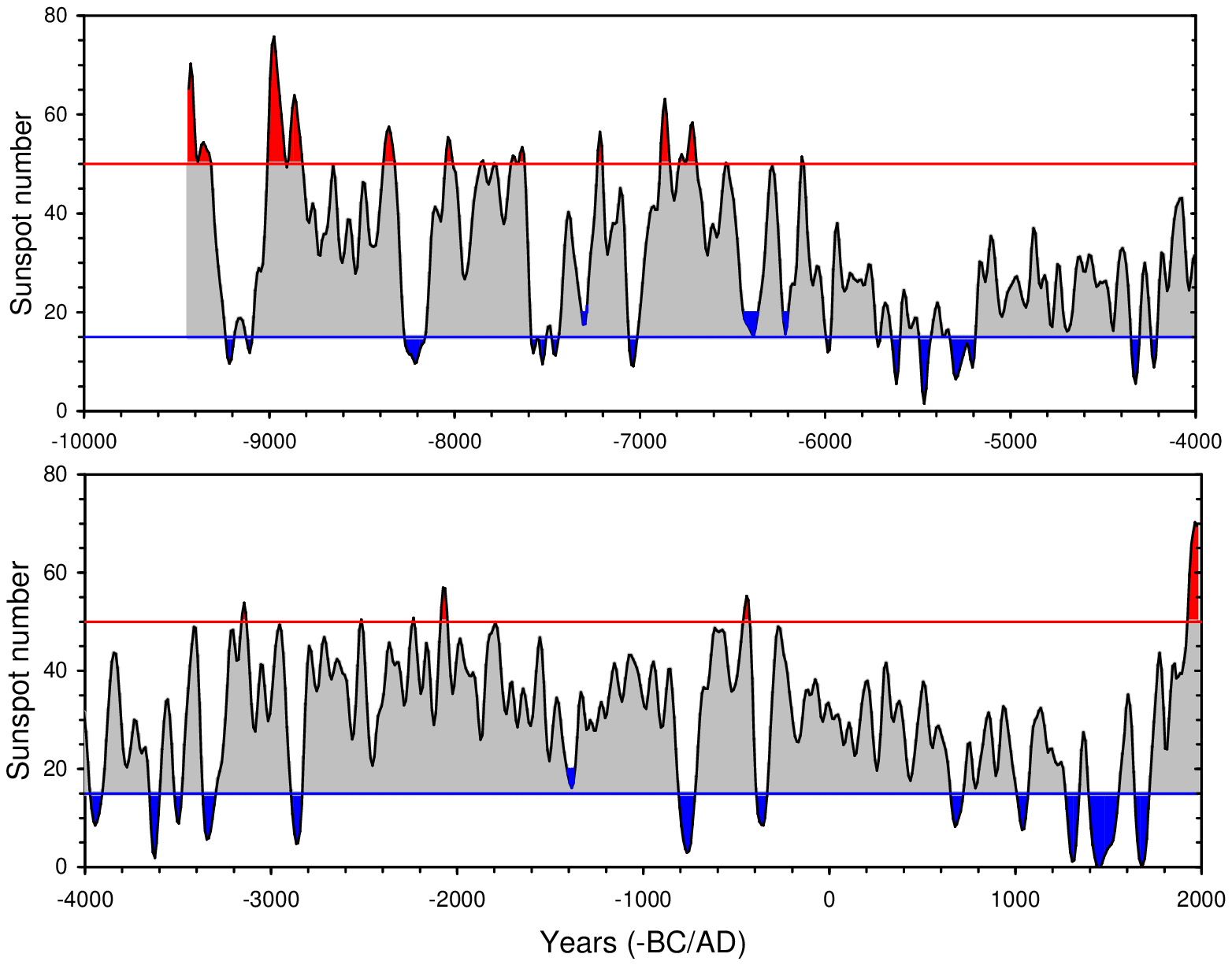}}
\caption{Sunspot number for the last 11400 years
reconstructed from 
$^{14}$C.
The original 10-year sampled data have been smoothed with a 1-2-2-2-1 filter
prior to plotting. Red and blue areas denote grand maxima and minima,
respectively.
Credit: \citet{usoskin-et-al-2007},
reproduced with permission \copyright ESO.
}
\label{Usoskinea07fig3}
\end{figure}

The reconstructed (smoothed) sunspot number is plotted in
Fig.~\ref{Usoskinea07fig3}.
Clearly, there have been a number of periods of very low activity similar to
the Maunder minimum. Those with the sunspot number
below 15 for at least 20 years are defined
as grand minima, while the sunspot number
above 50 for the same length of time gives a grand
maximum. Other definitions are also possible \citep[see,
e.g.,][]{abreu-et-al-2008}.

Grand minima and maxima are almost randomly distributed, although grand
minima tend to come in clusters separated by roughly 2000--3000 years.
Unlike the grand maxima, whose duration follows an exponential distribution,
the grand minima come in two varieties, a short (30--90 year) type, with
the Maunder minimum being a classic example, and a long ($>110$~year) type,
such as the Spoerer minimum (1390--1550).

The last grand maximum of solar activity has only just ended.
As \citet{usoskin-et-al-2003a} and \citet{solanki-et-al-2004} showed, the
Sun entered in a grand maximum in the middle of the 20th century,
characterized by strong sunspot cycles, short, comparitively active minima,
a high value of the Sun's open magnetic flux and plentiful other indicators
of vigorous solar activity.
This grand maximum has now ended as had been expected by
\citet{solanki-et-al-2004} and later by \citet{abreu-et-al-2008}.
This is indicated by the long and very quiet activity minimum between cycles
23 and 24 (2005--2010) and the weak currently running cycle.

Making predictions about how the activity will develop in future is not at
present possible beyond the maximum of the current cycle. Thus only
statistical estimates of future solar activity can be made based on
comparisons with the reconstructed long-term activity record
\citep{solanki-et-al-2004,abreu-et-al-2008,lockwood-2009,solanki-krivova-2011}.
In particular, it is unlikely that the Sun will slip into a grand minimum
\citep[less than 8\% likelihood within the next 30 years,][]{lockwood-2009}
and it is equally likely that the next grand extremum will be a grand
maximum as a grand minimum.


\subsection{By How Much Did the Sun Vary Between the Maunder Minimum
and Now?}
\label{secular}

\begin{figure}%
\epsfxsize35pc         %
\caption{Various TSI reconstructions since 1600 identified in the plot.
The dark blue vertical bar shows the possible range of the TSI change
following \citet[][see text, no
reconstruction available]{schrijver-et-al-2011}.
Other vertical bars denote uncertainties of the models, plotted in
same colours. Note that the uncertainty in the \citealt{shapiro-et-al-2011a}
model, $\pm3$~Wm$^{-2}$, extends downward outside the plot, and
the blue horizontal bar and arrow mark the reduced value of this model as
argued by \citet{judge-et-al-2012}.
The black dotted line shows the TSI value representing solar 
minimum conditions following SORCE/TIM measurements (see Sect.~\ref{obs}).
}
\centerline{\epsfbox{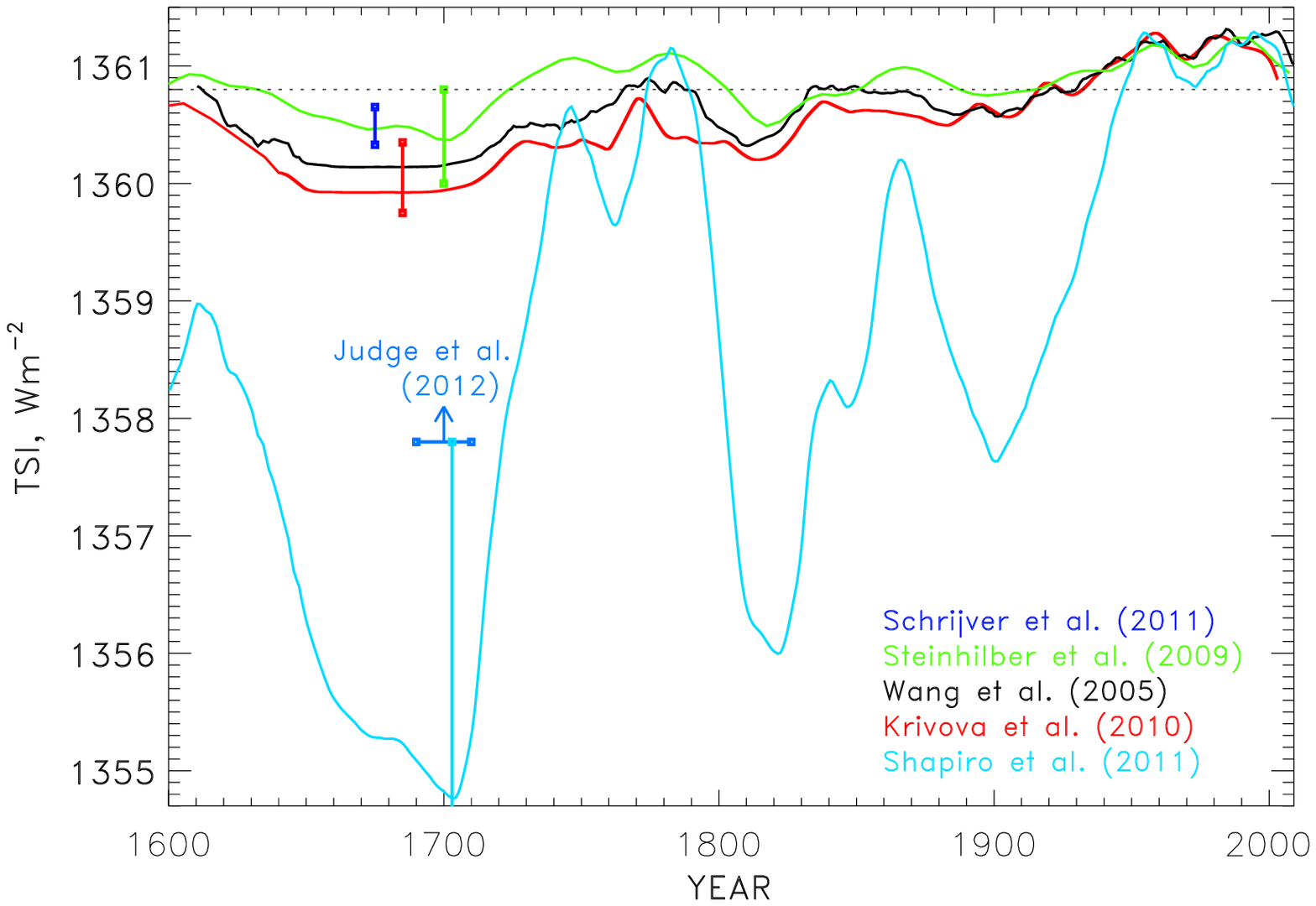}}
\label{tsi-1610}
\end{figure}

The sunspot number is a good representative of the solar magnetic activity
cycle.
Hence historical records of sunspot number since 1610
\citep{hoyt-schatten-98} and sunspot areas since 1874  \citep[e.g.,][and
references therein]{balmaceda-et-al-2009} allow decent reconstructions of
the cyclic component of solar irradiance changes over the last four
centuries.
All such reconstructions show that in the last few decades the Sun was
unusually active (see previous section), so that the cycle-average TSI was
also roughly 0.6~Wm$^{-2}$ higher than during the Maunder minimum
\citep{solanki-fligge-2000a} even in the absence of any secular change.

Reconstructions of the heliospheric magnetic field from the geomagnetic
aa-index and observations of the interplanetary magnetic field imply that
the Sun's open magnetic field increased by nearly a factor of two since the
end of the 19th century \citep{lockwood-et-al-99,lockwood-et-al-2009a} before
dropping again to the 19th century values in the last few years.
The total photospheric magnetic flux, which is more directly related to
solar irradiance, has been regularly measured for only about four decades
\citep[e.g.][]{arge-et-al-2002,wang-et-al-2006}, so that longer-term changes
cannot yet be reliably assessed.
\citet{harvey-93,harvey-94b} noticed, however, that small ephemeral active
regions keep bringing copious amounts of magnetic flux to the solar surface
during activity minima, when active regions are rare or absent.
The magnetic flux provided by the ephemeral regions, and concentrated in the
network in the quiet Sun, varies little over the activity cycle as ephemeral
regions belonging to two different solar cycles emerge in parallel for
multiple years
\citep{harvey-92,harvey-93,harvey-94b,hagenaar-et-al-2003}.
The overlap between the cycles provides a physical explanation for the
secular change in the photospheric magnetic field and irradiance
\citep{solanki-et-al-2000,solanki-et-al-2002}.

The magnitude of the secular change remains, however, heavily debated.
This is because the sunspot numbers or areas widely employed in the
reconstructions on time scales longer than a few decades are related only
indirectly to the amount of flux emerging in small ephemeral regions feeding
the magnetic network.

The first estimates of the TSI change since the Maunder minimum,
mainly derived from solar-stellar comparisons, ranged from
2 to 16~Wm$^{-2}$ \citep{lean-et-al-92,zhang-et-al-94,mendoza-97}.
They were indirect and based on a number of assumptions that were later found to be
spurious \citep[e.g.,][]{wright-2004,hall-lockwood-2004,hall-et-al-2009}.
\citet{judge-et-al-2012} argue that current stellar data do not yet allow
an assessment of the secular change in the solar brightness, and longer stellar
observations are required.



Various empirical reconstructions produced in the 2000s give values between
1.5 and 2.1~Wm$^{-2}$
\citep[e.g.,][]{foster-2004,lockwood-2005,mordvinov-et-al-2004}.
\citet{lockwood-stamper-99} were the first to apply a linear relationship
between the open magnetic flux and the TSI derived from the data obtained over
the satellite period.
Later, such a linear relationship was also employed by
\citet{steinhilber-et-al-2009} to reconstruct TSI from the $^{10}$Be data.
This reconstruction covers the whole Holocene and is discussed in
Sect.~\ref{holocene}, where we also consider 
the validity of the linear relationship.
Their reconstruction for the period after 1610 is shown in Fig.~\ref{tsi-1610}
together with a number of other recent reconstructions.
The derived TSI increase since 1710 is 0.9$\pm0.4$~Wm$^{-2}$.
Note that due to the uncertainties in the TSI levels during the last three
activity minima (see Sects.~\ref{obs} and \ref{modelling}) used to construct
the linear relationship, the uncertainty of this model is also relatively
high.

Models that are more physics-based were employed by \citet{wang-et-al-2005},
\citet{krivova-et-al-2007a} and \citet{krivova-et-al-2010a}.
\citet{wang-et-al-2005} used a surface flux transport simulation of the
evolution of the solar magnetic flux combined with the NRLSSI irradiance
model (see Sect.~\ref{modelling}).
\citet{krivova-et-al-2007a,krivova-et-al-2010a} have reconstructed the
evolution of the solar magnetic flux from the sunspot number with the
1D model of
\citet{solanki-et-al-2000,solanki-et-al-2002,vieira-solanki-2010} and then
used the SATIRE model (Sect.~\ref{modelling}) to reconstruct the irradiance.
This combination is called SATIRE-T (for telescopic era).
They found that the cycle-averaged TSI was about
1.3$^{+0.2}_{-0.4}$~Wm$^{-2}$ higher in the recent period compared with the
end of the 17th century, in agreement with the assessments by
\citet{foster-2004,lockwood-2005} and \citet{wang-et-al-2005}.
Both SATIRE-T and NRLSSI models are
shown in Fig.~\ref{tsi-1610}.

Most of the models are tested against the directly measured TSI and
reproduce it fairly well.
However, as the secular change over the satellite period is quite weak, if
any, and is not free of uncertainties, as indicated by the difference
between the three composites (see Sects.~\ref{obs} and \ref{modelling}, as
well as Figs.~\ref{fig-ball} and \ref{fig-ball-pmod}), these data are not
well suited to constrain the rise in TSI since the Maunder minimum.
For this reason the SATIRE model is also tested against other available data
sets.
Thus, the modelled solar total and open magnetic flux are
successfully compared to the
observations of the total magnetic flux over the last four decades and the
empirical reconstruction of the heliospheric magnetic flux from the aa-index
over the last century, respectively.
Also, the activity of the $^{44}$Ti isotope \citep{usoskin-et-al-2006b}
calculated from the SATIRE-T open flux agrees well with $^{44}$Ti activity
measured in stony meteorites \citep{vieira-et-al-2011}.
Finally, the reconstructed irradiance in Ly-$\alpha$ agrees with the
composite of measurements and proxy models by \citet{woods-et-al-2000a} going
back to 1947 \citep{krivova-et-al-2010a}.

Recently,
\citet{schrijver-et-al-2011} argued that the last minimum in 2008,
which was deeper and longer compared to the eight preceding minima, might
be considered as a good representative of a grand minimum.
This would mean a secular decrease of only about 0.15--0.5~Wm$^{-2}$
(estimated as the difference between TSI in the PMOD composite during the
minima preceding cycles 22 and 24 in 1986 and 2008, respectively,
\citealp{froehlich-2009}).
If the cycle averaged 0.6~Wm$^{-2}$ TSI change due to the cyclic component
\citep{solanki-fligge-2000a} is added to this, the total increase would be about
0.75--1.1~Wm$^{-2}$.
(Note that most sources list the sum of the modelled
cyclic and secular components of
the irradiance change since the Maunder minimum.)

A very different estimate was published by
\citet{shapiro-et-al-2011a}, who assumed that during the Maunder minimum the
entire solar surface was as dark as is currently observed only in the
dimmest parts of supregranule cells (in their interiors).
To describe this quiet state, \citet{shapiro-et-al-2011a} employed the
semi-empirical model atmosphere A by \citet[][see
Sect.~\ref{modelling}]{fontenla-et-al-99}, which gave a rather large TSI
increase of 6$\pm3$~Wm$^{-2}$ since the Maunder minimum.
This reconstruction is also shown in Fig.~\ref{tsi-1610}.
\citet{judge-et-al-2012} have recently argued, based on the analysis of
sub-mm data, that by adopting model~A \citet{shapiro-et-al-2011a}
overestimated quiet-Sun irradiance variation by about a factor of two, so
that the modelled increase in TSI since the Maunder minimum is overestimated
by the same factor.

\citet{foukal-milano-2001} argued on the basis of uncalibrated historic
Ca~II photographic plates from the Mt Wilson Observatory that the area
coverage by the network did not change over the 20th century, which would
imply no or very weak secular change in the irradiance.
A number of observatories around the globe carried out full-disc solar
observations in the Ca~II~K line since the beginning of the 20th century,
and some of these have recently been digitised.
\citet{ermolli-et-al-2009} have shown, however, that such historical images 
suffer from numerous problems and artefacts.
Moreover, calibration wedges are missing on most of the images, making proper
intensity calibration a real challenge.
Without properly addressing these issues, results based on historic images
must be treated with caution. 
In summary, present-day estimates of the TSI change since the end of the
Maunder minimum range from 0.8~Wm$^{-2}$ to about 3~Wm$^{-2}$, i.e. over
nearly a factor of 4. In addition, the time dependence also is different in
the various reconstructions and is rather uncertain.

\subsection{Variation over the Holocene}
\label{holocene}

In their 2004 review Fr{\"o}hlich and Lean \nocite{froehlich-lean-2004}
concluded that ``Uncertainties in
understanding the physical relationships between direct magnetic modulation
of solar radiative output and heliospheric modulation of cosmogenic proxies
preclude definitive historical irradiance estimates, as yet.'' Since then,
this topic has progressed rapidly and we now have several reconstructions
of TSI over the Holocene. These build upon the reconstructions of solar
activity indices (modulation potential, open flux, total flux, etc.)
described in Sect.~\ref{grand}, although 
only cycle averaged values of irradiance can be reconstructed prior to 1610. 

Since the modulation potential, $\Phi$, is the primary quantity obtained
from the production rates of cosmogenic isotopes (Sect.~\ref{grand})
it serves as an input to
all irradiance reconstructions on millennial time scales.
But the methods are different.
Thus \citet{shapiro-et-al-2011a} scale irradiance changes linearly with the
22-yr averaged $\Phi$ calculated by \citet{steinhilber-et-al-2009} from
$^{10}$Be data.
As discussed in Sect.~\ref{secular}, the magnitude of the secular change is
derived in this model from a comparison of the current Sun at activity
minimum conditions with the semi-empirical model atmosphere describing the
darkest parts of the intergranule cells, and the model shows a
significantly stronger variability compared to other models
(Fig.~\ref{tsi-1610}).

In fact, the relationship between irradiance and $\Phi$ is
not straightforward
\citep{usoskin-et-al-2002b,steinhilber-et-al-2009,vieira-et-al-2011}.
Therefore \citet{steinhilber-et-al-2009} and \citet{vieira-et-al-2011} first
employ physical models to calculate the solar open magnetic flux from
$\Phi$ 
\citep[see also][]{usoskin-et-al-2002b,usoskin-et-al-2003a,solanki-et-al-2004}.
TSI is then reconstructed in the model by \citet{steinhilber-et-al-2009}
through a linear empirical relationship between the directly measured open
magnetic flux and TSI during the three recent activity minima.
\citet{vieira-et-al-2011}, in contrast, apply a physical model
\citep{vieira-solanki-2010} to compute the sunspot number and the total
magnetic flux from the reconstructed open flux and to show that irradiance
is modulated by the magnetic flux from two consecutive cycles (which is not
so suprising, see Sect.~\ref{secular}). 
Thus irradiance can be represented by a linear combination of the $j$th and
$j$th+1 decadal values of the open flux.
This implies that although employment of a linear relationship between TSI
and the open flux is not justified physically, it might work reasonably on
time scales longer than several cycles.

\begin{figure}%
\epsfxsize25pc         %
\caption{{\bf (a)} TSI reconstruction since 9500~BC using the SATIRE-M
(blue) and SATIRE-T (red) models.
{\bf (b)} Enlargement of panel (a) for the last 3000~years.
The gray shading marks the uncertainty due to different
reconstructions of the geomagnetic field.
Credit: \citet{vieira-et-al-2011},
reproduced with permission \copyright ESO.
}
\centerline{\epsfbox{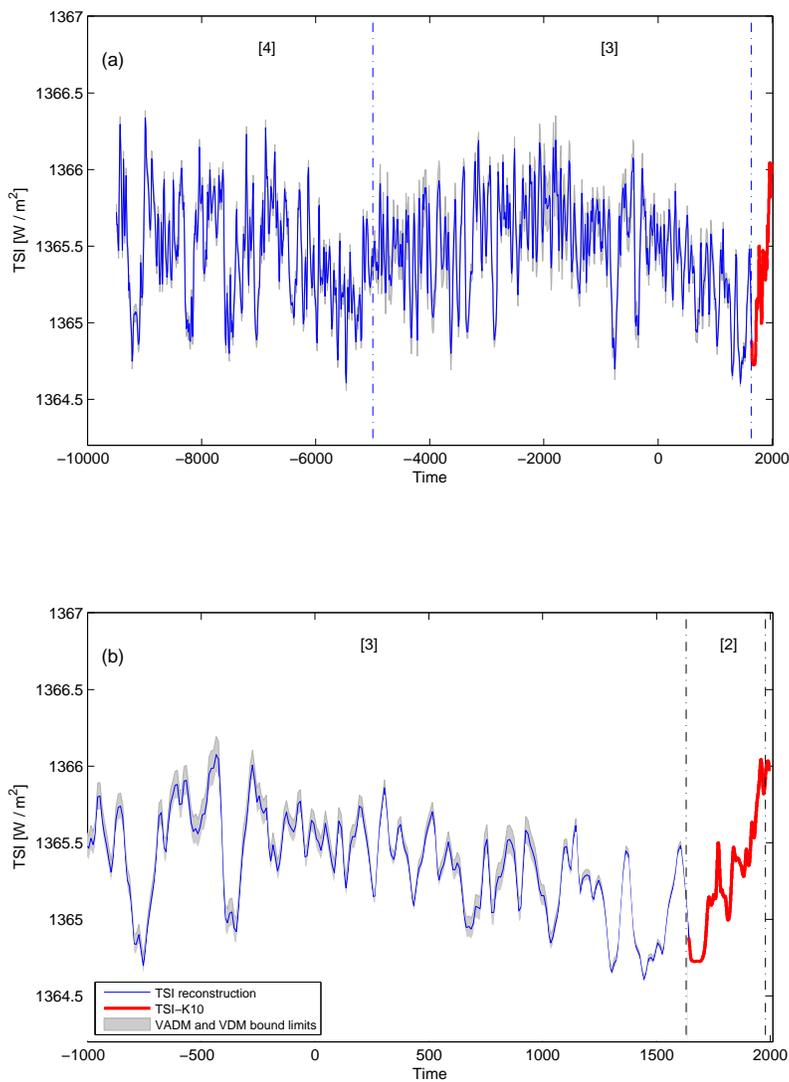}}
\label{tsi-holocene}
\end{figure}

The reconstruction of TSI over the Holocene by \citet{vieira-et-al-2011}
using the SATIRE-M (for Millennia) model is plotted in
Fig.~\ref{tsi-holocene}.
For comparison, the reconstruction from the telescopic sunspot record
(SATIRE-T, see previous section) is also shown.
In general, the various TSI reconstructions over the Holocene display
similar longer-term dependences \citep{steinhilber-et-al-2012}, with the
main difference being the amplitude of the variations (see
Sect.~\ref{secular}).
Thus, the reconstruction by \citet{steinhilber-et-al-2009} shows a somewhat
weaker variability, as can be judged from Fig.~\ref{tsi-1610} over the
telescopic era.

\section{INFLUENCE OF SOLAR VARIABILITY ON CLIMATE}
\label{climate}

\subsection{Evidence of Solar Influence on Climate on Different Time Scales}
\label{climate-evidence}

The role of the Sun in producing daily and seasonal fluctuations in
temperature, and their distribution over the Earth, seems so obvious that it
might be thought self-evident that variations in solar activity influence
weather and climate.  This idea has, however, been controversial over many
centuries.  The reasons for this scepticism centre around three areas:
firstly, the insubstantial nature of much of the meteorological ``evidence'';
secondly, a lack of adequate data on variations in solar energy reaching the
Earth and thirdly, related to the second, a lack of any plausible
explanation for how the proposed solar influence might take place.
Since the advent of Earth-orbiting satellites, however, we have substantial
evidence for variations in solar output, as discussed in
Section~\ref{short-term}, and
this, together with meteorological records of increasingly high quality and
coverage, are facilitating advances in understanding of solar signals in
climate.  In what follows we present some of the evidence for a solar
influence at the Earth's surface, and in the middle and lower atmospheres,
and go on to discuss the processes which might produce these signals.

\subsubsection{Surface.}
\label{surface}

Much work concerned with solar influences on climate has focussed on the
detection of solar signals in surface temperature.  It has frequently been
remarked that the Maunder Minimum in sunspot numbers in the second half of
the seventeenth century coincided with what is sometimes referred to as the
``Little Ice Age'' (LIA) during which most of the proxy records (indicators of
temperature including cosmogenic isotopes in tree rings, ice cores and
corals as well as documentary evidence) show cooler temperatures.  Care
needs to be taken in such interpretation as factors other than the Sun may
also have contributed.  The higher levels of volcanism prevalent during the
17th century, for example, would also have introduced a cooling tendency due
to a veil of particles injected into the stratosphere reflecting the Sun's
radiation back to space, this is discussed further in Section 4.2.1 below.

Evidence from a variety of sources, however, does suggest that during the
LIA the climate of the Northern Hemisphere was frequently characterised by
cooler than average temperatures in Eastern North America and Western
Europe, and warmer in Greenland and central Asia.  This pattern is typical
of a negative phase of a natural variation in climate referred to as the
North Atlantic Oscillation (NAO).  Indeed, temperature maps constructed from
a wide selection of proxy temperature data typically give the spatial
pattern of temperature difference between the Medieval Climate Anomaly (MCA,
c.950--1250) and the LIA, showing a pattern similar to a positive NAO.

Across the Holocene (the period of about 11,700 years since the last Ice
Age), isotope records from lake and marine sediments, glaciers and
stalagmites provide evidence that solar grand maxima/minima affect climate,
although these studies all rely on the reliability of the dating, which is
complex and not always precise.  The records show strong regional variations
typically including a NAO-like signal, as outlined above, and also a pattern
similar to a La Ni\~na event \footnote {This is the opposite phase of the
ENSO (El Ni\~no Southern Oscillation) cycle to El Ni\~no and is associated
with cold temperatures in the eastern Pacific Ocean)} and to greater monsoon
precipitation in southern Oman \citep[see e.g.  review
by][]{gray-et-al-2012}.

\begin{figure}
\epsfxsize22pc
\hspace*{-5mm}\centerline{\epsfbox{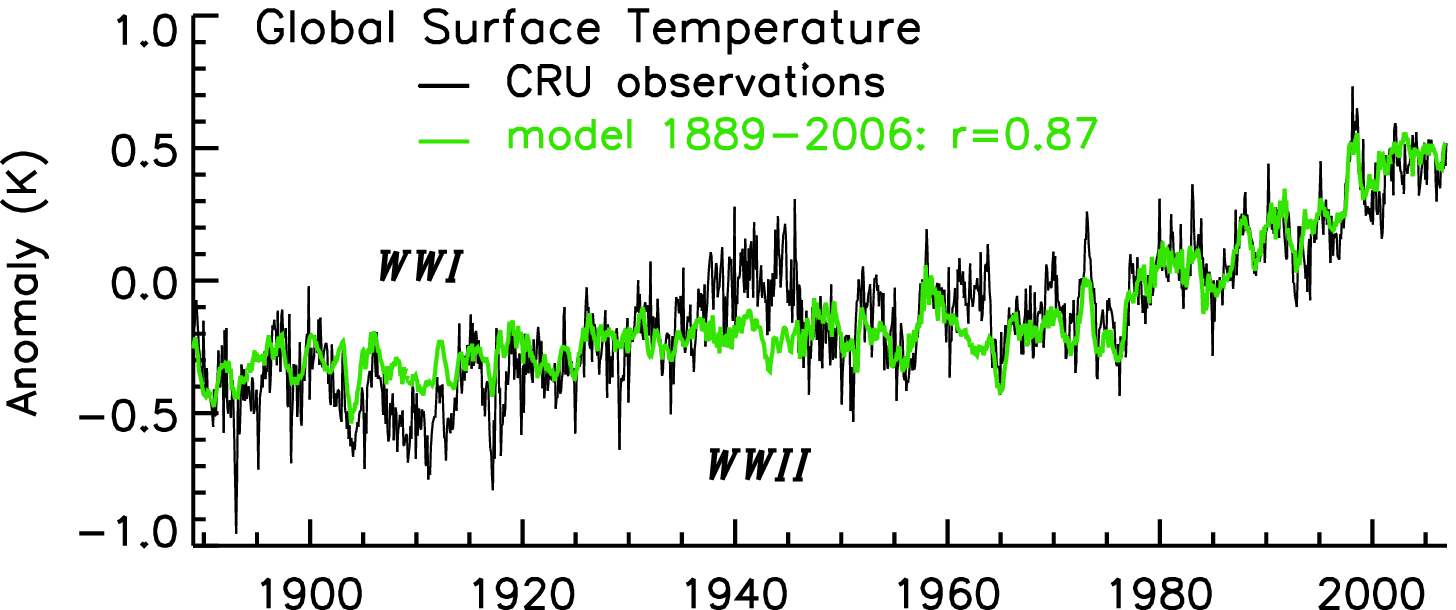}}
\epsfxsize25pc
\centerline{\epsfbox{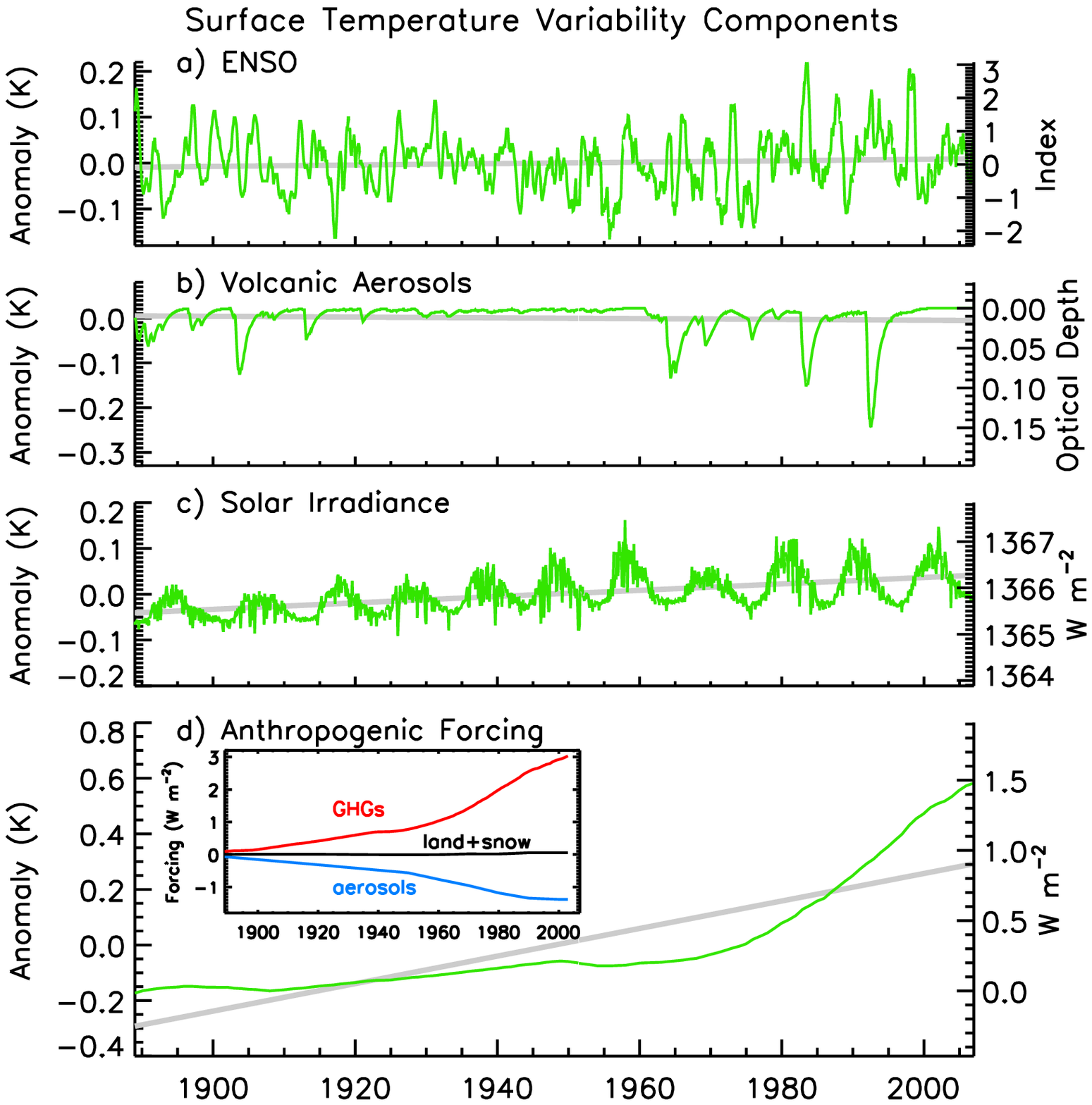}}
\caption{The top panel shows (in black) the global mean surface temperature
record compiled from measurements and (in green) its reconstruction from
a multiple linear regression model.  The lower four panels show the
contributions associated with the four regression components.  From
\citet{lean-rind-2008}}
\label{fig1}
\end{figure}

One approach, using a multiple linear regression analysis to separate
different factors contributing to global mean surface temperature over the
past century, is illustrated in Fig.~\ref{fig1}.  This suggests that the Sun
may
have introduced an overall global warming (disregarding the 11-year cycle
modulation) of approximately 0.07 K before about 1960, but that it has had
little effect since.  Over the century the temperature has increased by
about 1~K so the fractional contribution to global warming that can be
ascribed to the Sun over the last century is 7\%.  This result does,
however,
depend fundamentally on the assumed temporal variation of the solar forcing
and, as discussed in Section 3.2, there is some uncertainty in this.  The
index of solar variability used as the regression index in Fig.~\ref{fig1}
was that
of \citet{wang-et-al-2005}, which has a small long-term trend.  The effect on
radiative forcing of using different TSI records is further discussed in
Section 4.2.1

Crucially, however, it is not possible to reproduce the global warming of
recent decades using a solar index alone; this conclusion is confirmed by
studies using more sophisticated non-linear statistical techniques.

\begin{figure}
\epsfxsize30pc
\centerline{\epsfbox{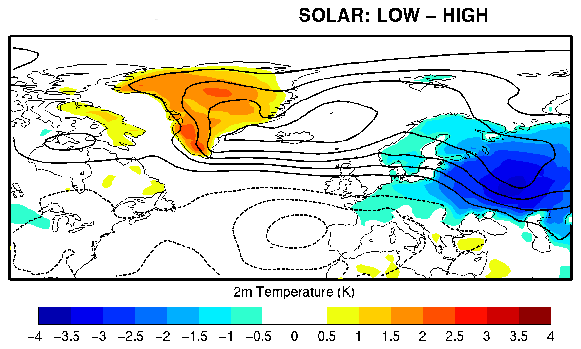}}
\caption{Difference in winter mean sea level pressure (contours, spaced at
1hPa with negative curves dashed) and near surface land temperature
(colours) between periods of low and high solar activity during the period
1957/8 to 2000/1.  From \citet{woollings-et-al-2010}}
\label{fig2}
\end{figure}

On the timescale of the 11-year solar cycle, analyses of surface temperature
and pressure show regional variations in the solar signal consistent with
those found over longer periods.  Figure~\ref{fig2}, for example, presents
the solar
cycle signal in the North Atlantic region derived from 44 winters of surface
temperature and pressure data \citep[from the ERA-40 Reanalysis dataset,
which
optimally combines observational and model data, see][]{uppala-et-al-2005}. 
The result is presented for periods of low, relative to high,
solar activity and resembles a negative NAO pattern.  This indicates that
atmospheric ``blocking events'', during which the jet-stream is diverted in a
quasi-stationary pattern associated with cold winters in Western Europe,
occur more frequently when the Sun is less active.

In the North Pacific Ocean \citet{christoforou-hameed-97} found that the
Aleutian Low pressure region shifts westwards when the Sun is more active
and the Hawaiian High northwards.  Solar signals in N.~Pacific mean sea
level pressure have also been identified, using different techniques, by
\citet{vanloon-et-al-2007} and \citet{roy-haigh-2010} implying shifts in
trade winds and storm tracks.

In the eastern tropical Pacific a solar cycle has been found by
\citet{meehl-et-al-2008} in sea surface temperatures (SST) which is
expressed as a cool
(La Ni\~na-like) anomaly at sunspot maximum, followed a year or two later by
a warm anomaly, although with data on only 14 solar peaks available the
robustness of this signal has been questioned by \citet{roy-haigh-2012}.  
Analyses of tropical circulations are not conclusive but a picture is
emerging of a slight expansion of the Hadley cells (within which air rises
in the tropics and sinks in the sub-tropics)
\citep[e.g.][]{broennimann-et-al-2007}, a strengthening of the Walker
circulation (an east-west circulation
with air rising over Indonesia and sinking over the eastern Pacific) and
strengthening of the South \citep{kodera-2004} and East
\citep{yu-et-al-2012} Asian monsoons when the Sun is more active.

Such changes in circulation are associated with changes in cloudiness, and
in the location and strength of regions of precipitation.  A solar signal in
cloudiness, however, remains difficult to establish, mainly due to the very
high innate variability in cloud and also to some uncertainty in the
definition of cloud types which might be affected.

\subsubsection{Atmosphere.}
\label{atmosphere}

Pioneering work in Berlin used data from meteorological balloons to show
correlations between stratospheric temperature and the solar 10.7 cm radio
flux over an increasing number of solar cycles \citep[see the summary
by][]{labitzke-2001}.  Largest correlations were found in the mid-latitude lower
stratosphere, implying temperature differences in that region of up to 1K
between minimum and maximum of the 11-year solar cycle.  This response was
very intriguing as it was much larger than would be expected based on
understanding of variations in irradiance.  Another interesting result was
the large warming found in the lower stratosphere at the winter pole.

Subsequently attempts to isolate the solar effect throughout the atmosphere
from other influencing factors have been carried out using multiple linear
regression analysis \citet{haigh-2003}, \citet{frame-gray-2010}.  In the
middle atmosphere the tropics show largest warming, of over 1.5K, in the
upper stratosphere near 1hPa, a minimum response around 5--30hPa and lobes of
warming in the sub-tropical lower stratosphere.  In the troposphere maximum
warming does not appear in the tropics but in mid-latitudes, with vertical
bands of temperature increase around 0.4K.

A similar analysis of zonal mean zonal (i.e. west-to-east) winds for the
Northern Hemisphere winter shows, when the Sun is more active, that there is
a strong positive zonal wind response in the winter hemisphere subtropical
lower mesosphere and upper stratosphere.  The zonal wind anomaly is observed
to propagate downwards with time over the course of the winter
\citep{kodera-kuroda-2002}.  In the troposphere the wind anomalies indicate
that the
mid-latitude jets are weaker and positioned further polewards when the Sun
is more active \citep{haigh-et-al-2005}.  This has implications for the
positions of the storm tracks and provides evidence for a solar signal in
mid-latitude climate.

It is well established that stratospheric ozone responds to solar activity. 
The vertically-integrated ozone column varies by 1--3\% in phase with the
11-year solar cycle, with the largest signal in the sub-tropics.  The
vertical distribution of the solar signal in ozone is more difficult to
establish, because of the short length of individual observational records
and problems of inter-calibration of the various instruments, but in the
tropics solar cycle variations appear to peak in the upper and lower
stratopshere with a smaller response between.

\subsection{Physical processes}
\label{processes}

\begin{figure}
\epsfxsize30pc
\centerline{\epsfbox{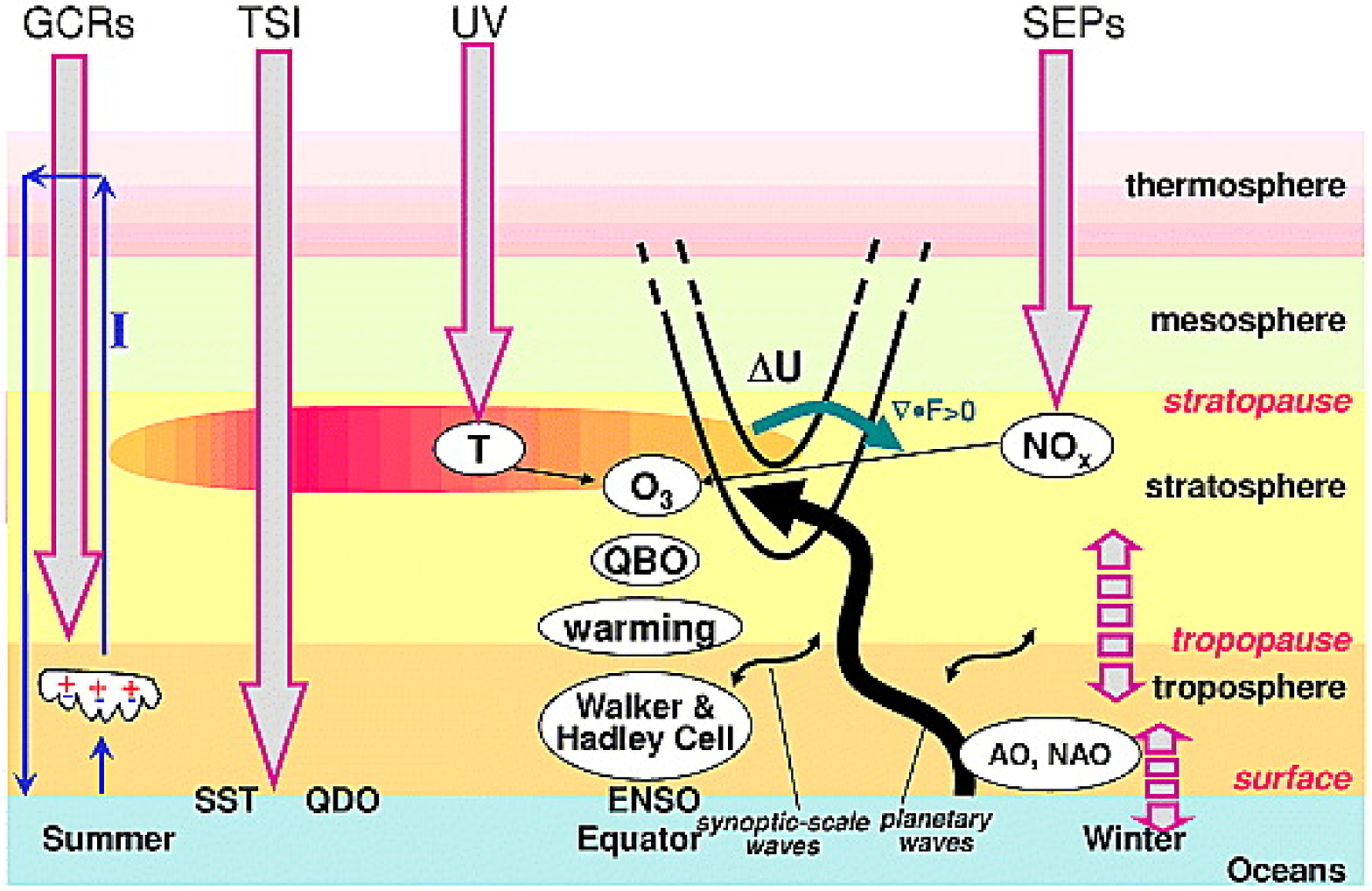}}
\caption{
Schematic indicating mechanisms whereby variations in solar activity may
influence the climate.  Solar changes are via total solar irradiance (TSI),
UV irradiance, solar energetic particles (SEPs) and galactic cosmic rays
(GCRs).  From \citet{gray-et-al-2012}}
\label{fig5}
\end{figure}

A range of physical and chemical processes, summarised in Fig.~\ref{fig5}, are
involved in producing the observed solar signals in climate, with some being
better characterised than others.  About one half of the total solar
irradiance entering the top of the atmosphere is transmitted to, and warms,
the Earth's surface so that variations in TSI have the potential to
influence climate through what have become known as ``bottom-up'' mechanisms. 
Solar ultraviolet radiation, however, is largely absorbed by the middle
atmosphere meaning that variations in UV have the potential to produce a
``top-down'' effect.  In either case the response of the atmosphere and oceans
involves complex feedbacks through changes in winds and circulations so that
the directive radiative effects provide only the initiating step.  It is
most likely that both routes for the solar influence play some role, further
increasing the overall complexity.

Also indicated in Fig.~\ref{fig5} are processes introduced through the effects of
energetic particles.  Galactic cosmic rays, the incidence of which is
modulated by solar activity, ionise the atmosphere, influencing the Earth's
magnetic field and possibly affecting cloud condensation.  Solar energetic
particles, emitted by the Sun during storms and flares, impact the
composition of the upper and middle atmosphere.  The means whereby either of
these effects may influence the climate (i.e.  mean temperature, wind,
precipitation) of the lower atmosphere are very uncertain.  In this article
we concentrate on radiative processes and do not offer any further
discussion of the effects of particles.

\subsubsection{Solar radiative forcing of climate.}
\label{RF}
 
The concept of Radiative Forcing (RF) is widely used \citep[see
e.g.][]{ipcc-2007}
in analysing and predicting the response of surface temperature to climate
change factors, including increasing concentrations of greenhouse gases,
higher atmospheric turbidity, changes in planetary albedo as well as changes
in solar input.  RF, in its most simple guise, is defined as the
(hypothetical) instantaneous change in net radiation balance produced at the
top of the atmosphere upon the introduction of a perturbing factor.  It is
useful because it has been shown (in experiments with general circulation
models (GCMs) of the coupled atmosphere-ocean system) that the change in
globally-averaged surface temperature, at equilibrium, is linearly related
to the RF value, and is much less dependent on the specifics of the forcing
factor.  The constant of proportionality, called the ``climate sensitivity
parameter'', $\lambda$, has a value estimated to be in the range 0.4--1.2
K~W$^{-1}$m$^2$ with a best estimate of 0.6~K~W$^{-1}$m$^2$
\citep[see e.g.][]{letreut-2012} indicating
that the equilibrated response of the global mean surface temperature to an
RF of 1~Wm$^{-2}$ would be 0.6~K.  We can use $\lambda$, together with
estimates of
time-varying TSI to indicate the role of the Sun in global climate change.  
It is important to note, however, that there is not a $1:1$ correspondence
between changes in TSI and RF.  This is because, while the Earth projects an
area of $\pi R^2$ to the Sun, the radiation is averaged over the $4\pi R^2$ of the
Earth's surface and, furthermore, only about 70\% of solar radiation is
absorbed with the other 30\% reflected back to space.  Thus a 1~Wm$^{-2}$
increase in TSI implies a RF of only $0.7/4 = 0.175$~Wm$^{-2}$ and (taking
$\lambda=0.6$~K~W$^{-1}$m$^2$) a global mean surface temperature increase of about
0.1~K.

A record of TSI may thus be used to indicate the role of the Sun in climate
history, at least in a global average equilibrated context.  A fundamental
issue then is to establish the TSI record and, as discussed in
Sect.~\ref{secular}, this is controversial.  As an example, plausible
estimates given for the
difference between the Maunder Minimum and the present probably lie in the
range 0.8--3.0~Wm$^{-2}$, suggesting a solar-driven global temperature
increase
in the range 0.08 to 0.30~K since the 17th century.  The observed
temperature difference is estimated at around 1~K so, on this basis, the Sun
may have contributed 8--30\% of the warming.

\begin{figure}
\epsfxsize20pc
\centerline{\epsfbox{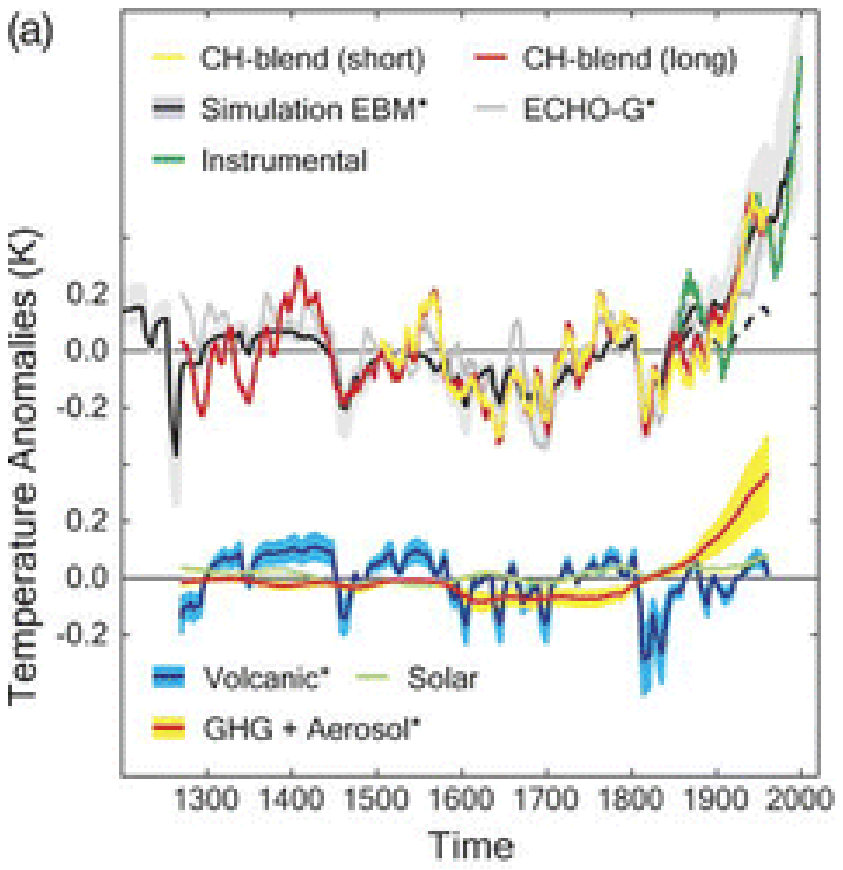}}
\caption{Top: Global mean surface temperature since 1270 constructed from
proxy data (red and yellow) and from instrumental records (green). 
Estimates of temperature calculated using an Energy Balance Model (black)
and a Global Climate Model (grey) and these are scaled to best match the
temperature reconstruction.
Bottom: An estimate of the contributions from individual forcings - volcanic
(blue), solar (green), and greenhouse gases and aerosols combined (red). 
From \citep{hegerl-et-al-2007}}
\label{fig6}
\end{figure}

In order to achieve a more accurate estimate it is useful to employ time series
of temperature record obtained from climate models with given external
forcings.  An example is presented for the last 800 years in
Fig.~\ref{fig6}.  This
shows the temperature record constructed from proxy and instrumental data
together with results derived from two climate models.  The first is a 2D
(horizontal with realistic distribution of land and ocean) Energy Balance
Model, which estimates the temperature of the atmosphere and upper ocean, in
response to time varying radiative forcings parameters, while taking account
of slower heat exchange with the deep ocean.  The second is a fully-coupled
atmosphere-ocean GCM.  The models are driven by changes in
greenhouse gases, tropospheric aerosol (sulphate, dust and soot particles),
stratospheric (volcanic) aerosol, as well as TSI.  The lower panel shows the
components of the temperature changes attributed to the individual forcings.  
The models suggest a TSI contribution at the low end of the range cited
above, consistent with the results of statistical analyses such as presented
in Fig.~\ref{fig1} for the 20th century.

We conclude that while solar activity, and volcanism, are very likely to
have contributed to variations in global (or hemispheric) average
temperature over the millennium, including to the LIA and MCA, they cannot
account for the sharp increase in warming since about 1960.

\subsubsection{A bottom-up mechanism for the influence of solar irradiance
variability on climate.}
\label{bottom-up}

Radiative forcing provides an indication of the global mean surface
temperature response to variations in TSI but cannot explain the regional
climate signals ascribed to solar variability, as outlined in Section 4.1. 
These cannot be driven directly by radiative processes, must be associated
with changes in atmospheric circulation, and so to understand them it is
necessary to look more deeply into the mechanisms involved.

The greatest intensity of solar radiation incident on Earth is in the
tropics but most reaches the surface in the cloud-free sub-tropical regions.  
Over the oceans a large proportion of this radiant energy is used in
evaporation.  The resulting high humidity air is advected into the tropics
where it converges and rises, producing the deep cloud and heavy
precipitation associated with that region.  The main ``bottom-up'' mechanism
for solar-climate links suggests that changes in the absorption of radiation
in the clear-sky regions provide the driver \citep{cubasch-et-al-97}.  
Greater irradiance would result in enhanced evaporation, moisture
convergence and precipitation.  This would result in stronger Hadley and
Walker circulations and stronger trade winds, driving greater upwelling in
the eastern tropical Pacific Ocean, colder SSTs and thus the La-Ni\~na-like
signal described in section 4.1.1 above.  There is some evidence for this
effect being reproduced in GCM simulations of solar effects
\citep{meehl-et-al-2008}, although the timing of the signal relative to the
solar cycle peak remains contentious \citep{roy-haigh-2012}.

\subsubsection{Solar spectral irradiance.}
\label{SSI}

The absorption of solar radiation by the atmosphere is determined by the
spectral properties of the component gases and is thus a strong function of
wavelength.

\begin{figure}
\epsfxsize30pc
\centerline{\epsfbox{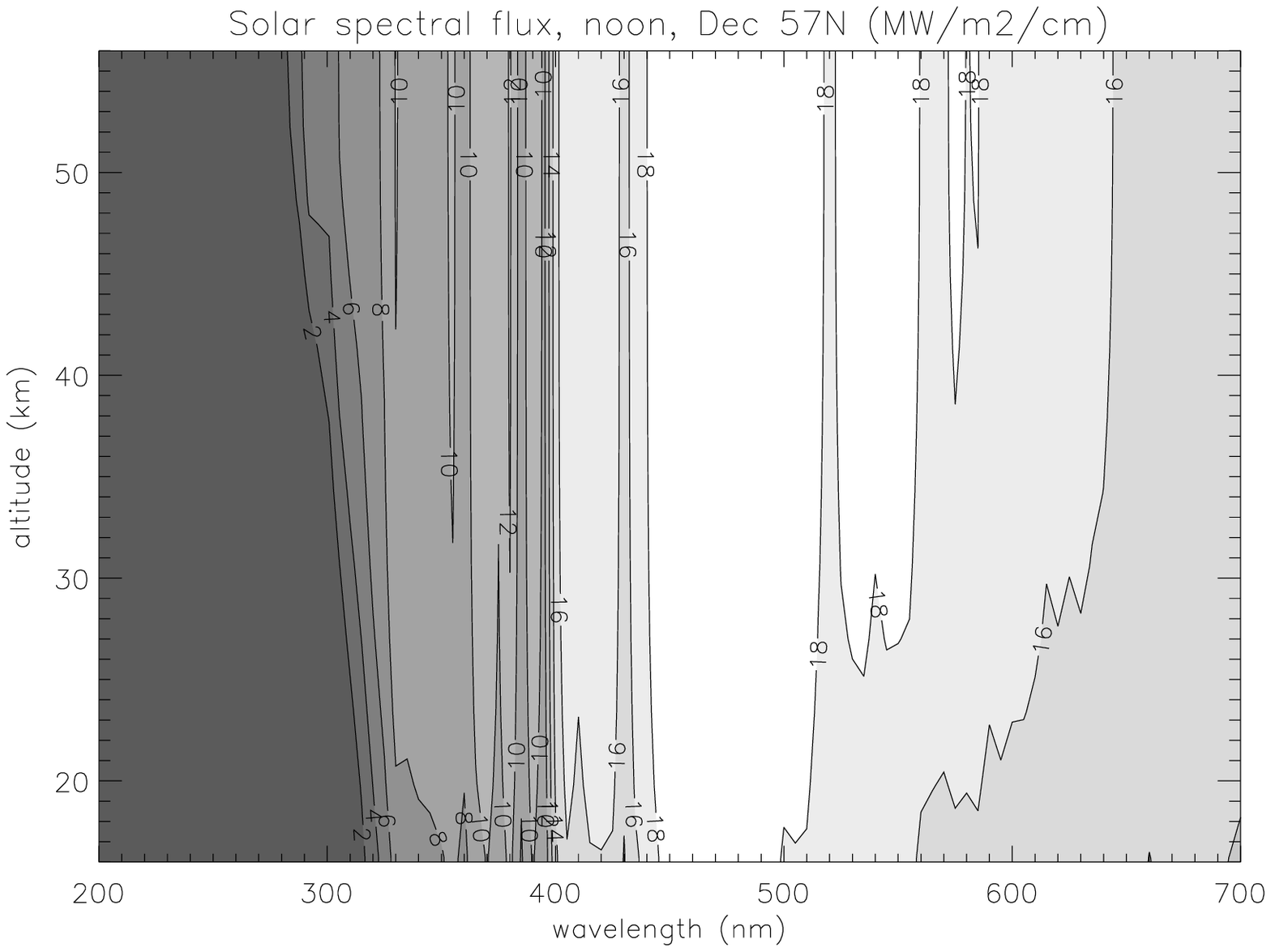}}
\epsfxsize30pc
\centerline{\epsfbox{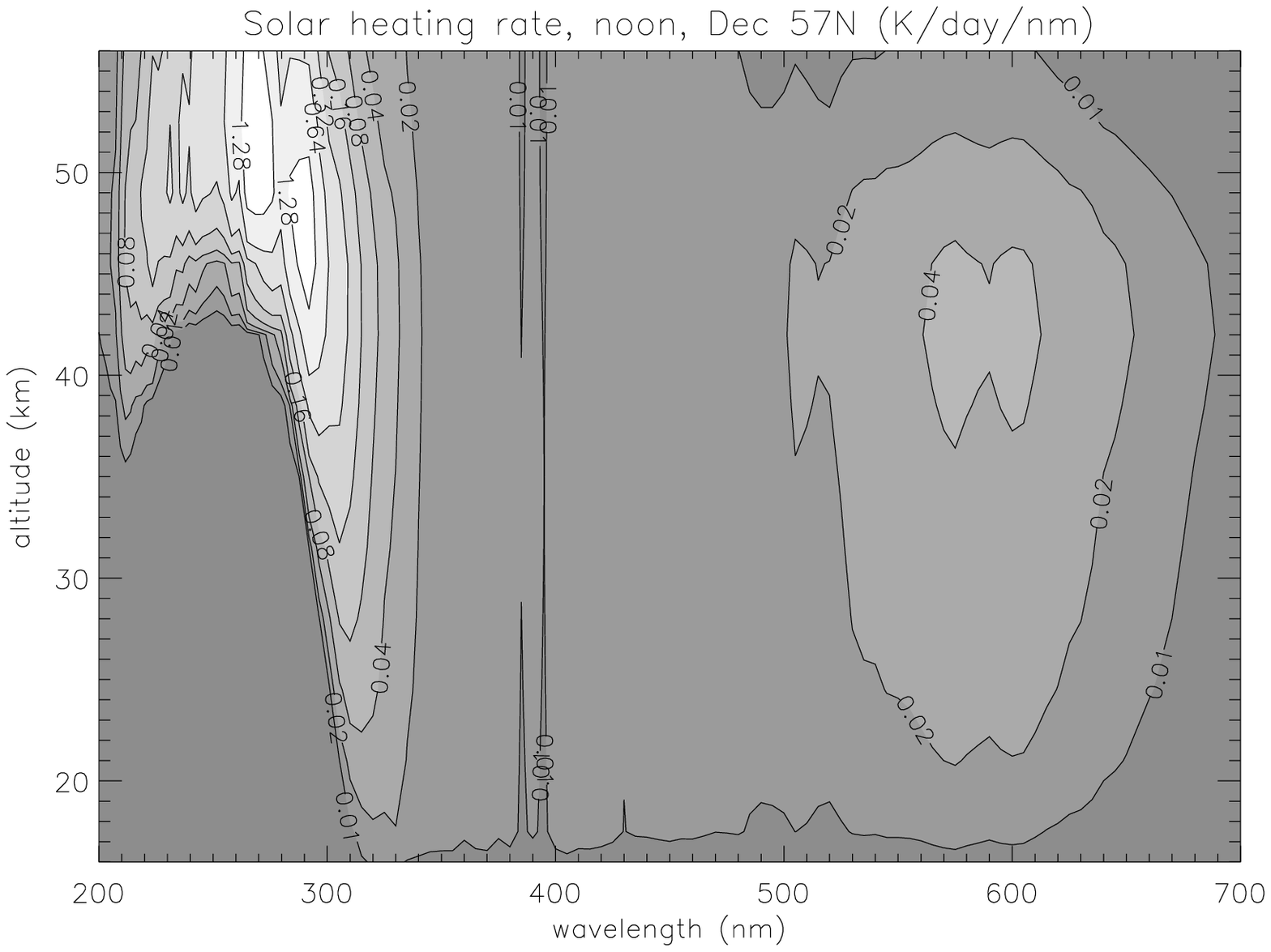}}
\caption{(a) Solar spectral irradiance (MWm$^{-2}$cm$^{-1}$) at UV and
visible wavelengths as a function of altitude within the Earth's atmosphere
calculated \citep[using a coupled radiative-chemical-dynamical 2D
model,][]{haigh-94} for latitude 57$^\circ$N, 21st December, noon, using
spectral irradiance
data at top of atmosphere for year 2000 from \citet{lean-2000}.
(b) Solar spectral heating rate (Kd$^{-1}$nm$^{-1}$).
}
\label{fig7}
\end{figure}

Figure~\ref{fig7}(a) shows an example of how downward solar spectral
irradiance, in
the near-UV and visible, depends on wavelength and on altitude within the
atmosphere.  As the radiation progresses downwards it is absorbed
preferentially at wavelengths shorter than 350 nm and longer than 440~nm;
Figure~\ref{fig7}(b) indicates how the absorption of radiation translates
into
atmospheric heating rates.  Absorption by molecular oxygen of radiation in
the 200--242nm region produces the oxygen atoms important in the production
of ozone and also heats the stratopause region.  Between 200 and 350 nm the
radiation is responsible for the photodissociation of ozone and for strong
radiative heating in the upper stratosphere and lower mesosphere.  The ozone
absorption bands, 440--800nm, are much weaker but, because they absorb
broadly across the peak of the solar spectrum, their energy deposition into
the lower stratosphere is not insignificant.

\begin{figure}
\epsfxsize30pc
\centerline{\epsfbox{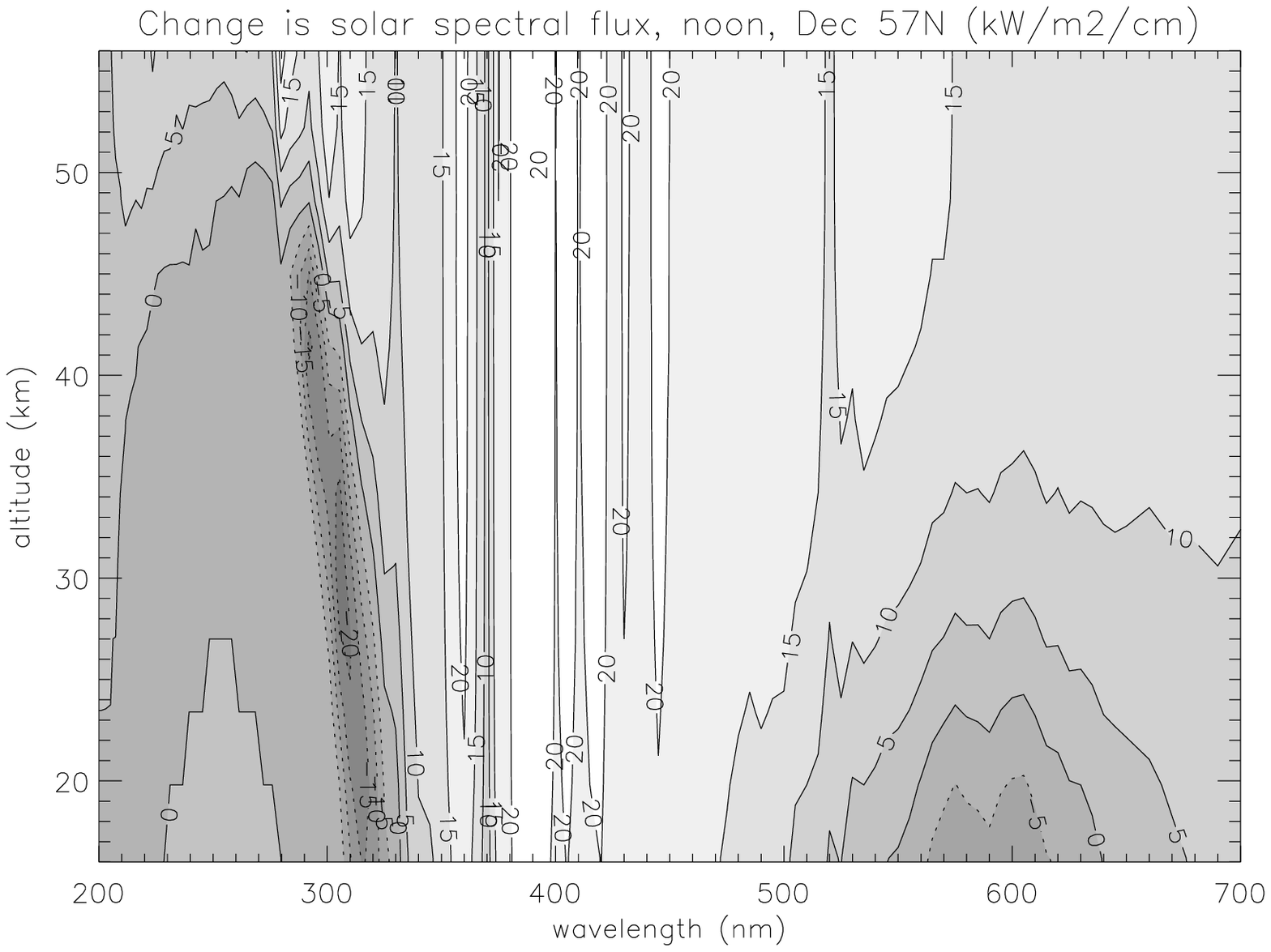}}
\caption{As in Figure 7(a) but showing the difference in spectral irradiance
(KWm$^{-2}$cm$^{-1}$) between 2007 and 2000 (representing the minimum and maximum of
the last  solar activity cycle).}
\label{fig8}
\end{figure}

Figure~\ref{fig8} shows the field of spectral irradiance, as in
Fig.~\ref{fig7}(a), but for
the difference between solar cycle maximum and minimum conditions, based on
spectral variability of \citet{lean-2000} (see Sect.~\ref{modelling}). 
In these plots the
effects of changes in ozone concentration resulting from the enhanced solar
irradiance are included.  This means that the vertical penetration of the
enhanced irradiance is not spectrally uniform because it depends on the
ozone perturbation as well as the ozone absorption spectrum.  Much of the
increased irradiance penetrates to the surface but at wavelengths less than
320 nm the increases in ozone result in lower values of irradiance
throughout the stratosphere, despite the increased insolation above.  In
similar fashion, at wavelengths in the 550--640 nm range, less radiation
reaches the lower stratosphere.  This effect is more marked at high solar
zenith angles, and so leads to strong latitudinal gradients in the
spectrally integrated irradiance in winter mid-latitudes.  This gives an
indication of the non-linearity of the response in solar heating rates to
variations in solar variability due to details of the photochemical response
in ozone.

Recent satellite measurements (see Sect.~\ref{obs}) suggest that solar UV
radiation varies by a much larger factor than assumed in Fig.~\ref{fig8}, while
even the sign of the change in radiation at visible wavelengths is uncertain
(\citealp{harder-et-al-2009}, cf. \citealp{ermolli-et-al-2012a}).
 These preliminary results significantly affect
estimates of the changes in heating rates, temperature and ozone fields and,
if such a spectral variability were confirmed to be representative of solar
cycle behaviour, or, indeed, of longer timescales, it would raise questions
concerning current understanding of the response to solar variability of the
temperature and composition of the middle atmosphere and also solar
radiative forcing of climate \citep{haigh-et-al-2010}.
However, there is some doubt about the reality of the extreme spectral
dependence of irradiance variations found by \citet{harder-et-al-2009};
see \citet{deland-cebula-2012,ermolli-et-al-2012a}
and Sects.~\ref{obs} and \ref{modelling}.

\subsubsection{Dynamical effects in the stratosphere.}
\label{strat dyn}

    Attempts to predict the response of the stratosphere to solar UV
variability were first carried out using 2-D chemistry transport models,
such as used to estimate the solar radiation fields in Fig.~\ref{fig7} and
Fig.~\ref{fig8}.  
These predicted a solar cycle response with a peak warming of around 1K
near the stratopause and peak increases in ozone of around 2\% at altitudes
around 40~km, with perturbations in both temperature and ozone monotonically
decreasing towards the tropopause \citep[e.g.,][]{garcia-et-al-84,haigh-94}.  
They did not reproduce the more complex latitudinal and vertical gradients,
the double peak structure in the strtaosphere described above.  This
indicates that the
ozone response, at least in the middle and lower stratosphere, is influenced
by modifications to its transport brought about by solar-induced changes in
atmospheric circulation.  Furthermore, as the structure in the temperature
signal is fundamentally related to the ozone response it is unlikely that
any simulation will satisfactorily reproduce the one without the other.

It has long been appreciated that variations in solar heating affect the
dynamical structure of the middle atmosphere.  Changes in the meridional
temperature gradients influence the zonal wind structure and thus the
upwards propagation of the planetary-scale waves which deposit momentum and
drive the mean overturning circulation of the stratosphere.  The changed
wind structure then has a further effect on wave propagation, as represented
in the schematic of Fig.~\ref{fig5}.  With greater solar heating near the
tropical
stratopause the stratospheric jets are stronger, the polar vortices less
disturbed and the overturning circulation weaker \citep{kodera-kuroda-2002}.  
This produces cooling in the polar lower stratosphere due to weaker descent,
and warming at low latitudes through weaker ascent.  Thus in the lower
stratosphere the tropics are warmer, and the poles cooler, than would result
from radiative processes alone.

Models with fully interactive chemistry have been employed so that the
imposed irradiance variations affect both the radiative heating and the
ozone photolysis rates, allowing feedback between heating, circulation and
composition \citep[see the review by][]{austin-et-al-2008h}.  These models
are broadly able to simulate the observed vertical structure of the solar
signal in ozone in the tropics, although there is no clear picture as to
what factor is responsible for the lower stratospheric maximum.
Candidates
include time-varying sea surface temperatures, transient solar input, high
vertical resolution in the models and their ability to produce natural modes of
variability such as the Quasi-Biennial Oscillation or ENSO.

\subsubsection{A top-down mechanism for the influence of solar irradiance
variability on climate.}
\label{top-down}

\begin{figure}
\epsfxsize26pc
\hspace*{5mm}\centerline{\epsfbox{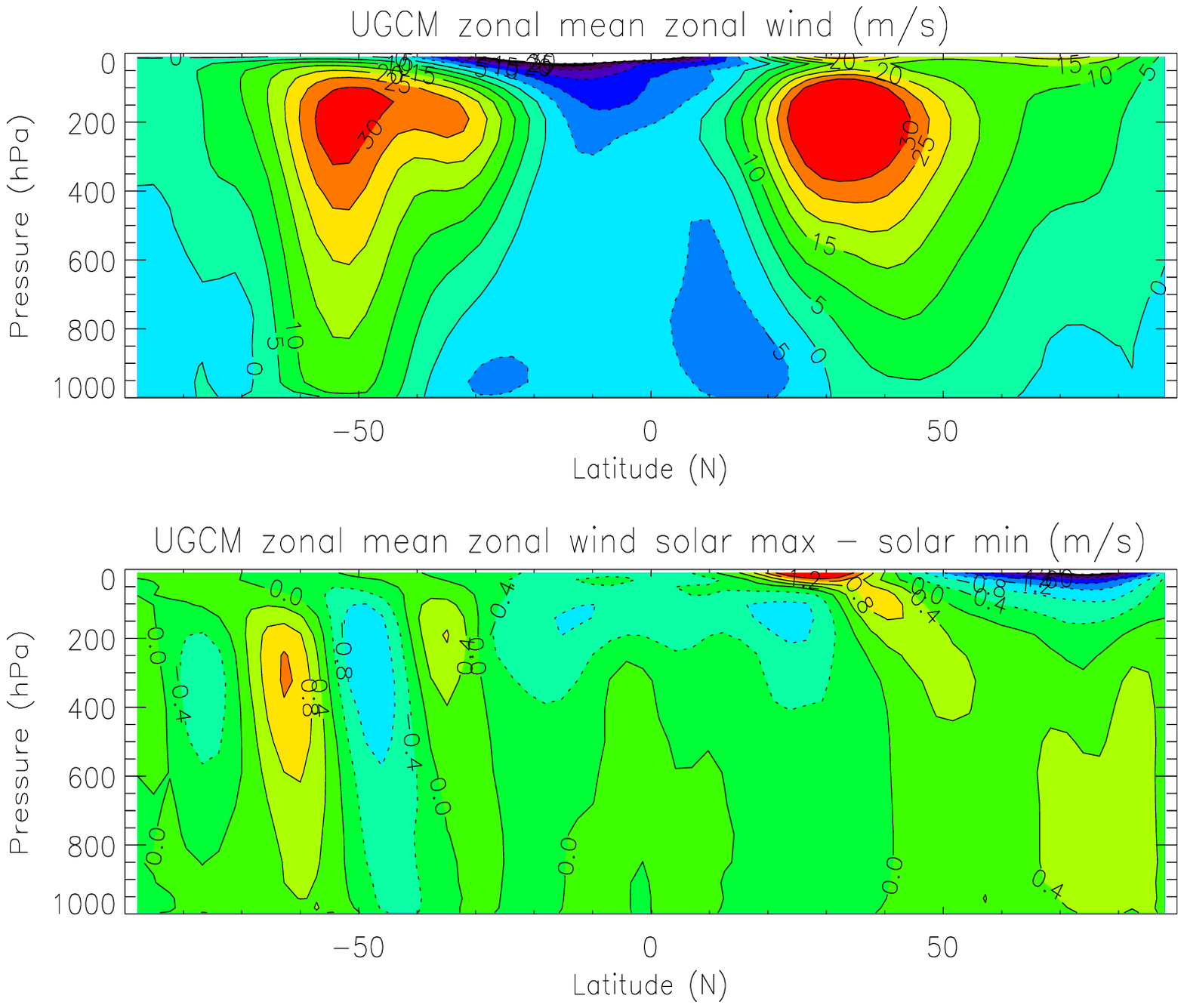}}
\epsfxsize30pc
\centerline{\epsfbox{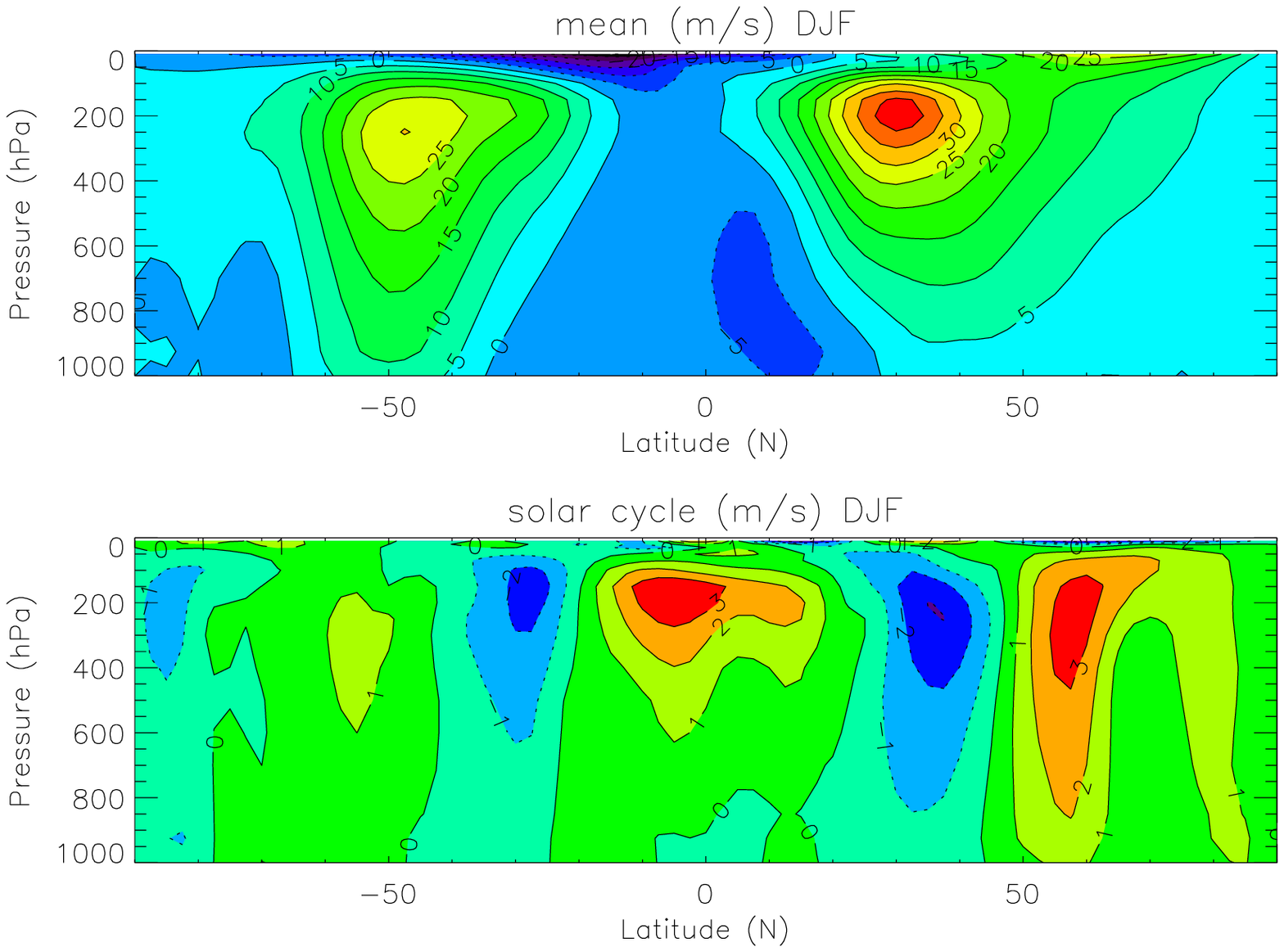}}
\caption{1st panel: Annual and zonal mean zonal wind (ms-1) as a function of
latitude and pressure altitude from a climate model.  2nd: Difference in
zonal mean zonal wind between solar max and min of the 11-year cycle
calculated by imposing changes in UV in the model  \citep{haigh-99}. 3rd: As
first panel but data from the NCEP dataset.  4th: Solar signal (max-min) in
NCEP dataset derived using multiple linear regression \citep{haigh-2003}.}
\label{fig9}
\end{figure}

While the bottom-up mechanism is entirely plausible, it is premised on very
small fractional variations in TSI, and also awaits confirmation of details
of the processes involved.  The larger fractional changes in the UV
radiation suggest an alternative route.  Studies of the impact of varying UV
in climate models in which sea surface temperatures have been fixed, are at
least qualitatively successful in simulating the tropospheric patterns of
response to solar variability \citep{haigh-96,haigh-99,larkin-et-al-2000,%
matthes-et-al-2006,shindell-et-al-99}.  For example, in
Fig.~\ref{fig9}, the lower two panels show the zonal wind climatology and the solar
cycle signal, respectively from a multiple regression analysis of the NCEP
Reanalysis dataset (http://www.cdc.noaa.gov/).  The solar signal appears as
an inverted horseshoe pattern in the troposphere, representing the poleward
shift of the jets at solar maximum, as mentioned in Section 4.1.2.  The upper
two panels present the same fields calculated with an atmosphere-only GCM in
response to increasing solar UV radiation.  In both analyses the jets weaken
and move polewards and both also produce (not shown) the broadened Hadley
cells.  The strength of the model signal is weaker than that seen in the
observations but is strengthened when larger increases in stratospheric
ozone are imposed.  Consistent with this, the Hadley circulation response in
coupled chemistry simulations \citep{shindell-et-al-2006} have been linked to
the additional heating introduced into the tropical upper troposphere and
lower stratosphere by solar-induced ozone.  The model studies clearly reveal
a dynamical influence of changes in the stratosphere on the troposphere
rather than a direct radiative effect.

More generally, a number of different studies have indicated that such a
downward influence does take place.  Analyses of observational data suggest
a downward propagation of polar circulation anomalies in both the Northern
and Southern hemispheres.  Model studies have also demonstrated a downward
influence from Antarctic stratospheric ozone depletion on the circumpolar
circulation in the southern hemisphere \citep{gillett-thompson-2003} and
from stratospheric temperature trends on the NAO \citep{scaife-et-al-2005}. 
These studies did not specifically address the impact of solar variability
on climate, but they did suggest that the troposphere responds to
perturbations initiated in the stratosphere.  More recently, a study by
\citet{ineson-et-al-2011}, using a coupled atmosphere-ocean GCM, prescribed
changes to solar UV as suggested by SORCE-SIM measurements (in the
200--320~nm range only) and produced a significant shift to a negative NAO
pattern, and colder winters in Western Europe, at lower solar activity~---
as suggested by observational records and presented in Fig.~\ref{fig2}.

There are many mechanisms proposed whereby the lower stratosphere may exert
a dynamical influence on the troposphere
\citep[see reviews by][]{shepherd-2002,haynes-2005,gerber-et-al-2012}. 
These include a response of the
mean meridional circulation to angular momentum forcing from above,
modification of the transmission of upward propagating planetary-scale
waves, and feedbacks between changes in the mean-flow and tropospheric
baroclinic eddies (weather systems).  

Coupling between the Hadley circulation and the mid-latitude eddies may also
play a key part.  Studies with a simple climate model
\citep{haigh-et-al-2005}, in which an anomalous heating was applied in the
tropical lower stratosphere, found a zonal mean tropospheric response
qualitatively similar to that observed in response to the solar cycle.  

Further experiments have investigated the chain of causality involved in
converting the stratospheric thermal forcing to a tropospheric climate
signal \citep{haigh-blackburn-2006,simpson-et-al-2009}.
They found
that changes to the thermal structure of the lower stratosphere influenced
the propagation of synoptic scale waves, creating anomalies in eddy heat and
momentum fluxes which drove changes in zonal wind and meridional circulation
throughout the troposphere.  These tropospheric changes then influenced the
subsequent propagation of waves so as to reinforce the initial
perturbations.  They concluded that solar heating of the stratosphere may
produce changes in the circulation of the troposphere even without any
direct forcing below the tropopause, and that the impact of the
stratospheric changes on wave propagation is key to this effect.

Although details of the mechanisms involved are still not fully established
it is becoming increasingly clear that solar variability may influence the
climate of the troposphere through processes whereby UV heating of the
stratosphere indirectly influences the troposphere through dynamical
coupling.


\section{CONCLUSIONS}
\label{conclusions}


Space-based radiometers have recorded total solar irradiance (TSI) since
1978 and have established that it varies at the 0.1\% level over the solar
cycle.
Solar variability is a
strong function of wavelength, increasing towards shorter wavelengths and
thus reaching a factor of two in the Ly-$\alpha$ line.
Most of the irradiance variability of the Sun is produced by dark (sunspots,
pores) and bright (magnetic elements forming faculae and the network)
surface magnetic features, whose concentration changes over the solar cycle.

Various radiometers show many similarities in their results. Nonetheless, a
few important differences are present that need to be removed before the
solar influence on climate can be accurately estimated.
One particularly relevant open issue is how strongly TSI changes on time
scales longer than the solar cycle.
Composites of TSI records from different instruments put
together by different scientists display a broad variety of behaviours. 
Models of TSI show the best agreement with the so-called PMOD composite.
In spite of this general concensus between models, the model-based estimates
of the rise in TSI since the Maunder minimum differ by almost a factor of four,
with correspondingly different effects on the outputs of climate
models.  Aside from this scaling issue, the models do agree relatively well
(e.g.  in their temporal behaviour), so that the irradiance can now be
reconstructed over the whole Holocene, covering multiple grand minima and
maxima of solar activity.

Another bone of contention is the behaviour of the spectral irradiance over
the solar cycle.
The modelled spectral irradiance at  all wavelengths, except around the
opacity minimum region in the infrared, varies in phase with the solar cycle
and with the TSI.
The SIM instrument on the SORCE satellite, by contrast, finds an antiphase
behaviour over large parts of the visible wavelength range and a UV
variability larger by a factor of 2--6 than that found by any model that also
reproduces TSI.  There are hints that the problem may lie with the SORCE SSI
data.

There is growing evidence that changes in solar irradiance affect the
Earth's middle and lower atmosphere.
At higher levels of activity the stratosphere is warmer throughout the
tropics, associated with higher concentrations of ozone.
At such times patterns consistent with an expansion of the tropical Hadley
circulations are established.
At the surface a more active Sun is associated with a more positive phase of
North Atlantic Oscillation, especially in winter, and a north-westwards
shift of the main surface pressure features in the North Pacific.

Globally the mean surface temperature varies in phase with solar activity. 
Over the past few solar cycles, for which measurements of TSI are available,
the small amplitude of this variation is consistent with what would be
deduced from radiative forcing arguments (of order 0.1K for a 1~Wm$^{-2}$
change in total solar irradiance).  On longer timescales, while the in-phase
relationship broadly persists back into the past, the amplitude is less easy
to associate directly with radiative forcing because of the large
uncertainties in TSI variations.  However, it is virtually impossible to
assign the global warming of the past half century to variations in solar
irradiance alone, using either statistical or physical methods.

In understanding the distribution of the solar signals throughout the
troposphere and across the Earth's surface, it is necessary to invoke
processes other than direct radiative heating alone.  The most plausible
scenarios involve changes in circulation, winds etc.  in response to
absorption of visible radiation at the surface and/or ultra-violet radiation
in the stratosphere.
Two main mechanisms have been proposed: the bottom-up mechanism relying on
absorption of mainly visible radiation at the surface and the top-down
mechanism relying on absorption of UV radiation in the stratosphere. 
The first of these is dependent
on enhanced heating and evaporation at the cloud-free sub-tropical ocean
surface with impacts on the inter-tropical convergence zone and tropical
circulations in the atmosphere and ocean.  The second involves changes in
the thermal structure of the stratosphere with dynamical coupling downwards.
At present the theory of, and evidence for, the stratospheric route is better
developed, although uncertainty in
solar UV variability leaves scope for this to be revised.
The two routes are not mutually exclusive and may operate
synergistically.

In summary, considerable progress has been made in the last decade on the
topic of solar irradiance variability and its influence on climate, but new
data and models have also revealed new inconsistencies that will provide a
challenge for the future. It is an interdisciplinary field that has
reverberations well beyond astrophysics and considerable effort will be
required to overcome the challanges ahead.




\paragraph{FUTURE ISSUES}

Further advances in understanding solar  variability and its influence on
climate will benefit from
continuing acquisition of high quality measurements of solar and climate
variables and the development of models including
all the relevant processes.
The next generation of solar irradiance
models will make use of model atmospheres obtained
from three-dimensional MHD simulations allowing a more realistic
representation of the spectral properties of magnetic features.

\paragraph{ACKNOWLEDGMENTS}

We thank G.~Kopp, W.~Ball, Y.~C.~Unruh,
I.~Usoskin for providing figures \ref{tsi-kopp},
\ref{fig-ball} \& \ref{fig-ball-pmod}, \ref{fig-yvonne},
\ref{Usoskinea07fig3}, respectively.
We are grateful to J.~Lean, A.~Shapiro and F.~Steinhilber for providing or
making public their TSI reconstructions (used in Fig.~\ref{tsi-1610}), as
well as to L.~Floyd, J.~Harder, M.~Snow and T.~Woods for the spectral
irradiance data (shown in Fig.~\ref{fig-ssi-var}).
JDH would like to acknowledge the many colleagues who contributed to the
Solar Influences on Climate Consortium funded by the UK Natural Environment
Research Council.
This work has been partially supported by the WCU grant (No.~R31--10016)
funded by the Korean Ministry of Education, Science and Technology.


\newcommand{\ApJ}{Astrophys.~J.}
\newcommand{\ApJL}{Astrophys. J. Lett.}
\newcommand{\ApJS}{Astrophys. J. Suppl. Ser.}
\newcommand{\AAp}{Astron. Astrophys.}
\newcommand{\AApR}{A\&AR}
\newcommand{\AApSS}{A\&AS}
\newcommand{\ARAA}{Annu. Rev. Astron. Astrophys.}
\newcommand{\ApSS}{Ap\&SS}
\newcommand{\AZh}{AZh}
\newcommand{\Afz}{Afz}
\newcommand{\AJ}{Astron.~J.}
\newcommand{\ApJSS}{ApJS}
\newcommand{\BAAS}{BAAS}
\newcommand{\EPS}{Earth, Planets and Space}
\newcommand{\JGR}{J. Geophys. Res.}
\newcommand{\GRL}{Geophys. Res. Lett.}
\newcommand{\MNRAS}{Mon. Not. R. Astron. Soc.}
\newcommand{\PASJ}{PASJ}
\newcommand{\PASP}{Publ. Astron. Soc. Pac.}
\newcommand{\PSS}{Planet. Space Sci.}
\newcommand{\QJRAS}{Q. J. R. Astron. Soc.}
\newcommand{\SPh}{Solar Phys.}
\newcommand{\CelMech}{Celest. Mech.}
\newcommand{\LRSP}{Liv. Rev. Sol. Phys.}

\bibliographystyle{Astronomy}



\paragraph{ACRONYMS AND DEFINITIONS}


\begin{description}
\item
ENSO~--- El Ni\~no Southern Oscillation
\item
GCMs~--- General Circulation Models of the coupled atmosphere-ocean system
\item
Grand minimum (maximum)~--- multi-decade period of low (high) solar activity
\item
LIA~--- Little Ice Age; a period of cooling in the Northern Hemisphere
(coldest part around 17th century)
\item
Maunder minimum~--- the most famous example of a grand minimum (17th
century)
\item
MCA~--- the Medieval Climate Anomaly (also Medieval Warm Period or Medieval
Climate Optimum); a period (c.950--1250) of warm climate in the North
Atlantic region
\item
NAO~--- North Atlantic Oscillation; a pattern of climate fluctuations, which 
controls the strength and direction of
westerly winds across the North Atlantic
\item
NRLSSI~--- the Naval Research Laboratory Solar Spectral Irradiance model
widely used in climate simulations
\item
RF~--- Radiative Forcing, the
(hypothetical) instantaneous change in net radiation balance produced at the
top of the atmosphere upon the introduction of a perturbing factor
\item
SATIRE~--- a model for Spectral And Total Irradiance
REconstructions
developed at Max-Planck-Institut f\"ur Sonnensystemforschung (MPS)
\item
SSI~--- Spectral Solar Irradiance
\item
SST~--- sea surface temperature
\item
TSI~--- Total Solar Irradiance; the total power from
the Sun impinging on a unit area (perpendicular to the Sun's rays) at 1AU
(given in units of Wm$^{-2}$), integral over SSI
\end{description}

\end{document}